\documentclass[12pt,preprint]{aastex}

\slugcomment{Accepted for publication in ApJ}

\shorttitle{Emergence of a BAL Outflow in WPVS 007}
\shortauthors{Leighly et al.}

\begin{document}


\title{Emergence of a Broad-Absorption-Line Outflow in the Narrow-line
  Seyfert 1 Galaxy WPVS 007\footnote{Based on observations made with
  the NASA-CNES-CSA Far-Ultraviolet Spectroscopic Explorer. FUSE is
  operated for NASA by Johns Hopkins University under NASA contract
  NAS5-32985.}}


\author{Karen M. Leighly\altaffilmark{1}}
\affil{Homer L. Dodge Department of Physics and Astronomy, The
  University of Oklahoma, 440 W.\ Brooks St., Norman, OK 73019}
\altaffiltext{1}{Visiting Professor, Fall 2006 -- Spring 2007, The
  Ohio State University,   Department of Astronomy, 4055 McPherson
  Laboratory, 140 West 18th   Avenue, Columbus, OH 43210-1173}
\email{leighly@nhn.ou.edu}

\author{Fred Hamann}
\affil{Department of Astronomy, University of Florida, 211 Bryant
  Space Science Center, Gainesville, FL 32611-2055}

\author{Darrin A. Casebeer}
\affil{Homer L. Dodge Department of Physics and Astronomy, The
  University of Oklahoma, 440 W.\ Brooks St., Norman, OK 73019}

\and

\author{Dirk Grupe}
\affil{Department of Astronomy and Astrophysics, Pennsylvania State
  University, 525 Davey Lab, University Park, PA 16802}



\begin{abstract}
We report results from a 2003 {\it FUSE} observation, and reanalysis
of a 1996 {\it HST} observation of the unusual X-ray transient
Narrow-line Seyfert 1 galaxy WPVS~007. The {\it HST} FOS spectrum
revealed mini-BALs with $V_{max}\sim 900\rm \, km\, s^{-1}$ and $FWHM
\sim 550 \rm \, km\, s^{-1}$.  The {\it FUSE} spectrum showed that an
additional BAL outflow with $V_{max} \sim 6000 \rm \, km\, s^{-1}$
and $FWHM \sim 3400 \rm \, km\, s^{-1}$ had appeared.  WPVS~007 is a
low-luminosity object in which such a high velocity outflow is not
expected; therefore, it is an outlier on the $M_V$/$V_{max}$
relationship.  Template spectral fitting yielded apparent ionic
columns, and a {\it Cloudy} analysis showed that the presence of
\ion{P}{5} requires a high ionization parameter $\log(U) \geq 0$ and
high column density $\log(N_H) \geq 23$ assuming solar abundances
and a nominal SED for low-luminosity NLS1s with
$\alpha_{ox}=-1.28$.  A recent long {\it Swift} observation revealed
the first hard X-ray detection and an intrinsic
(unabsorbed) $\alpha_{ox}\approx -1.9$. Using this SED in our
analysis yielded lower column density constraints
($\log(N_H) \geq 22.2$ for $Z=1$, or $\log(N_H) \geq 21.6$
if $Z=5$). The X-ray weak continuum, combined with X-ray absorption
consistent with the UV lines, provides the best explanation for the
observed {\it Swift} X-ray spectrum. The large column densities and
velocities implied by the UV data in any of these scenarios could be
problematic for radiative acceleration. We also point out
that since the observed \ion{P}{5} absorption can be explained
by lower total column densities using an intrinsically X-ray weak
spectrum, we might expect to find \ion{P}{5} absorption
preferentially more often (or stronger) in quasars that are
intrinsically X-ray weak.

\end{abstract}





\section{Introduction}

Active galaxies, the most luminous persistently-emitting objects in
the Universe, are powered by mass accretion onto a supermassive black
hole.  But gas not only falls into the black hole, it can also be
blown out of the central engine in powerful winds, as indicated by the
blue-shifted emission and absorption lines observed in the 
rest-frame UV spectra.  Outflows are likely to be an essential part of
the AGN phenomenon because they can carry away angular momentum and
thus facilitate accretion through the disk.  Winds are important probes of
the chemical abundances in AGN, which appear to be elevated
\citep{hf99}. They can distribute chemically-enriched gas through the 
intergalactic medium \citep{clm02}.  They may carry kinetic
energy to the host galaxy, influencing its evolution, and contributing
to the coevolution of black holes and galaxies implied by the observed
correlation between the black hole and bulge masses (e.g., Scannapieco
\& Oh 2004).

Unfortunately, the nature and origin of AGN outflows remain largely
mysterious. Blueshifted absorption lines are the most easily
identifiable evidence for outflows.  The absorption lines forming in
AGN winds are divided into three categories based on line widths:
broad absorption lines (BALs; FWHM $\sim 10,000 \rm \, km\, s^{-1}$),
narrow absorption lines (NALs; FWHM$\sim <500 \rm\, km \,s^{-1}$), and
intermediate mini-broad absorption lines (mini-BALs).  The
relationship between these types of outflows is not well understood,
although it appears that BAL outflows are seen only in luminous
quasars \citep[e.g.,][]{lb02}, with narrower lines observed in less
luminous objects.  It is widely believed that the outflows arise from
the accretion disk, but the point of origin and acceleration
mechanism(s) are not understood \citep[e.g.,][]{proga07}.  In
addition, BAL flows are associated with X-ray weakness
\citep[e.g.,][]{gallagher06}, suggesting absorption of the X-ray
continuum by very high columns approaching $10^{23} \,\rm cm^{-2}$,
that, if also outflowing, strongly constrain acceleration mechanisms.

There is a strong need to measure basic parameters of the flows, but 
it is difficult.  The projected velocity of the BAL is measured directly
from the absorption line profile, but the amount of matter and
kinetic energy in the outflow is difficult to constrain.  Lines are
saturated, although not black, implying that the absorbing  
material only partially covers the source. The geometry and radial
extent of the flows are not known. The density is difficult to
constrain because most lines arise from permitted transitions
and are not very density dependent. 

Line variability provides a valuable tool for studying the absorbing
gas.  For example, variation in absorption-line apparent optical
depths can sometimes be attributed to differences in ionization, and
the variability time scale can be used to place lower limits on the
electron density.  This information can then be used to constrain the
distance from the nucleus \citep[e.g.,][]{narayanan04}.  In addition,
line variability confirms the intrinsic nature of an outflow
\citep{rh07}.  But variations appear to be small in general; mini-BALs
become stronger or weaker, and portions of BAL profiles become
stronger or weaker.  For example, \citet{barlow94} found that 15 of 23
BALQSOs spectroscopically monitored showed variations in the BAL
ranging from marginal to large, but these variations were all limited
to changes in the residual intensity in the trough rather than
variations in the velocity structure.  \citet{lundgren07} studied a
sample of 29 BALQSOs that had been observed more than once during the
SDSS.  They found that 16 were significantly variable, but while they
found a few that showed significant variations in velocity structure,
there was again a trend for a greater incidence of variability in
residual trough intensity.  \citet{gibson08} report results of a study
of variability of 13 BALQSQs observed first in the Large Bright Quasar
Survey, and again in the SDSS, yielding a 3--6 year baseline in
rest-frame years.  All of the quasars vary, but in complex ways,
generally appearing as changes in trough depth in discrete regions a
few thousand kilometers per second wide.  Capellupo et al.\ (in prep.)
also find relatively modest variability to be common in a sample of
24 BALQSOs observed in multiple epochs.  All together, these studies
reveal a lack of dramatic variability indicating the flows are
generally stable and long term in typical BALQSOs.

WPVS~007 is a low-luminosity Seyfert 1 galaxy ($\alpha_{2000}$=00 39
15.2; $\delta_{2000}=-$51 17 02; $z=0.02882$, $M_V=-18.8$), that is
known to exhibit rather peculiar behavior.  It was observed 1990 Nov
10--12 during the {\it ROSAT} All Sky Survey (RASS), where it was
found to be a bright X-ray source, although it was observed to have
the softest spectrum ever  found in an AGN \citep[inferred
  $\alpha=-7.3$ for $F_E   \propto E^{\alpha}$;][]{grupe95}. Subsequent
observations (from 1993 Nov 11--13 to the present {\it Swift}
monitoring campaign) have found the  object to be
consistently X-ray weak \citep{grupe95, grupe07, grupe08}, and the
origin of the X-ray weakness is not known. Recently, a {\it Swift}
observation revealed the first detection of this object in hard X-rays
($>2 \rm \, keV$).  Although the number of photons observed were
small, the data suggest that the source is partially covered by an
absorber with $N_H \sim 10^{23}\rm \, cm^{-2}$ \citep{grupe08}.  

The UV observations presented in this paper provide a possible
explanation of the X-ray transience.  We observed dramatic variability
in the absorption line properties between an archival {\it HST}
spectrum from an observation in 1996 (discussed in \S 2) in which the  
object displays mini-BALs with a maximum velocity of about $900 \rm \,
km\, s^{-1}$ and $FWHM \sim 550 \rm \, km\, s^{-1}$, and a {\it FUSE}
spectrum from an observation in 2003 (\S 3) in which the object was
found to have developed broad absorption lines with $V_{max} \sim
6,000 \rm \, km\, s^{-1}$ and $FWHM \sim 3400 \rm \, km\, s^{-1}$ in
addition to the mini-BALs. This transformation is the most dramatic
ever seen in a active galaxy, and we postulate that the changing
absorption observed  in the UV is coupled to the  transient X-ray
behavior. We measure lower limits on the column densities of various
ions for both the BAL and the miniBAL (\S 2 \& \S 3), and use {\it
  Cloudy} models to estimate the physical parameters of the absorbing
gas.    A summary of the paper is given in \S 5, and appendices
describe details of the {\it FUSE} data analysis.   Unless  otherwise
specified, we assume $H_0=73\rm \, km\, s^{-1}\, Mpc^{-1}$,
$\Omega_M=0.27$, and a flat Universe.

\section{{\it HST} Observations}

The {\it HST} observations were conducted 1996 July 30.
The {\it HST} spectra are more than 10 years old, and the data have
appeared in the literature several times.  \citet{goodrich2000} shows
the spectrum in the region of Ly$\alpha$ and comments on the
absorption.  \citet{crenshaw99} include the data in a study of
absorption lines in Seyfert 1 galaxies.  They report absorption lines
from Ly$\alpha$, \ion{N}{5}, \ion{Si}{4}, and \ion{C}{4}.
\citet{cs03} analyze some of the properties of the UV spectra of
NLS1s, and construct a composite NLS1 spectrum of all publically
available {\it HST} spectra of NLS1s at the time of writing, including 
WPVS~007.  \citet{kura04} analyze all active galaxies observed with
the {\it HST} FOS after the COSTAR correction optics had been put in
place.   They perform an automated spectral fitting over a broad band
pass and present extensive tables of results.   However, while this
object is included in their paper, their tables do not include
information on the high-ionization lines at wavelengths from Ly$\alpha$ to
\ion{He}{2}~$\lambda 1640$.  \citet{dunn06} present light curves from
{\it IUE} and {\it HST}  observations of Seyfert galaxies. They
find evidence that WVPS~007 varied among four epochs of observation,
becoming faintest in the {\it HST} observation.  Finally,
\citet{grupe07} present {\it Swift} photometry\footnote{The central
  wavelengths of the {\it Swift} filters are as follows: $v$ 5468
  \AA\/; $b$ 4392 \AA\/; $u$ 3465 \AA\/; $uvw1$ 2600\AA\/; $uvm2$ 2246
  \AA\/; $uvw2$ 1928\AA\/ \citep{poole08}}, showing that WPVS~007
has brightened since the {\it HST} observation. 

Since none of the previously published results present the information
we need to do the required analysis of the absorption lines, we
extract the spectra from the archive and perform the measurements
ourselves. 

\subsection{Preliminary Analysis}

The spectra were extracted from the {\it HST} Data ``Online''
website\footnote{http://archive.stsci.edu/hstonline/} for Legacy
instruments including the {\it FOS}.  These are final calibration data
products, so no recalibration was necessary.  The observation log is
presented in Table 1.   

We first verify the wavelength calibration of the spectra by
measuring the wavelengths of Galactic absorption lines.  We used
\ion{Si}{2}~$\lambda 1260.3$,  \ion{C}{2}~$\lambda 1334.5$,
\ion{Al}{3}~$\lambda 1670.8$, \ion{Fe}{2} lines at 2344.2, 2382.8,
2586.6, and 2600.2\AA\/, and \ion{Mg}{2}~$\lambda\lambda 2796.4,
2803.5$ (note that vacuum wavelengths are used, as appropriate for the
{\it HST} data).  These lines showed that the spectral segments
obtained using the G130H, G190H, and G270H gratings were consistent
with no anomalous wavelength shifts.  There were no convenient
Galactic absorption lines available to examine the spectra obtained
with the G400H and G570H gratings; since we find no anomalous
redshifts or flux offsets (see below) we assume that these spectra are
also free of wavelength shifts. 

We next merge the spectra starting with the short wavelengths.  We
examined overlapping regions for flux offsets and found no evidence for
any. The final signal-to-noise ratio of the continuum varies from
$\sim 3.7$ at wavelengths shortward of 1600\AA\/ (observed frame), to
22 in near 2700\AA\/, and falling to $\sim 9$ around 6600\AA\/.  We smooth
the spectra using a three-point scheme in which the center point is
weighted three times the adjacent points.  We correct for the Galactic
reddening of $E(B-V)=0.012$ mag \citep{schegel98} using the
\citet{cardelli89} reddening curve.  Finally, we shift the spectrum
into the rest frame, adopting the NASA/IPAC Extragalactic Database
redshift value of  0.02882.  

\subsection{Continuum Shape and Reddening}

\begin{figure}
\figurenum{1}
\epsscale{1.0}
\plotone{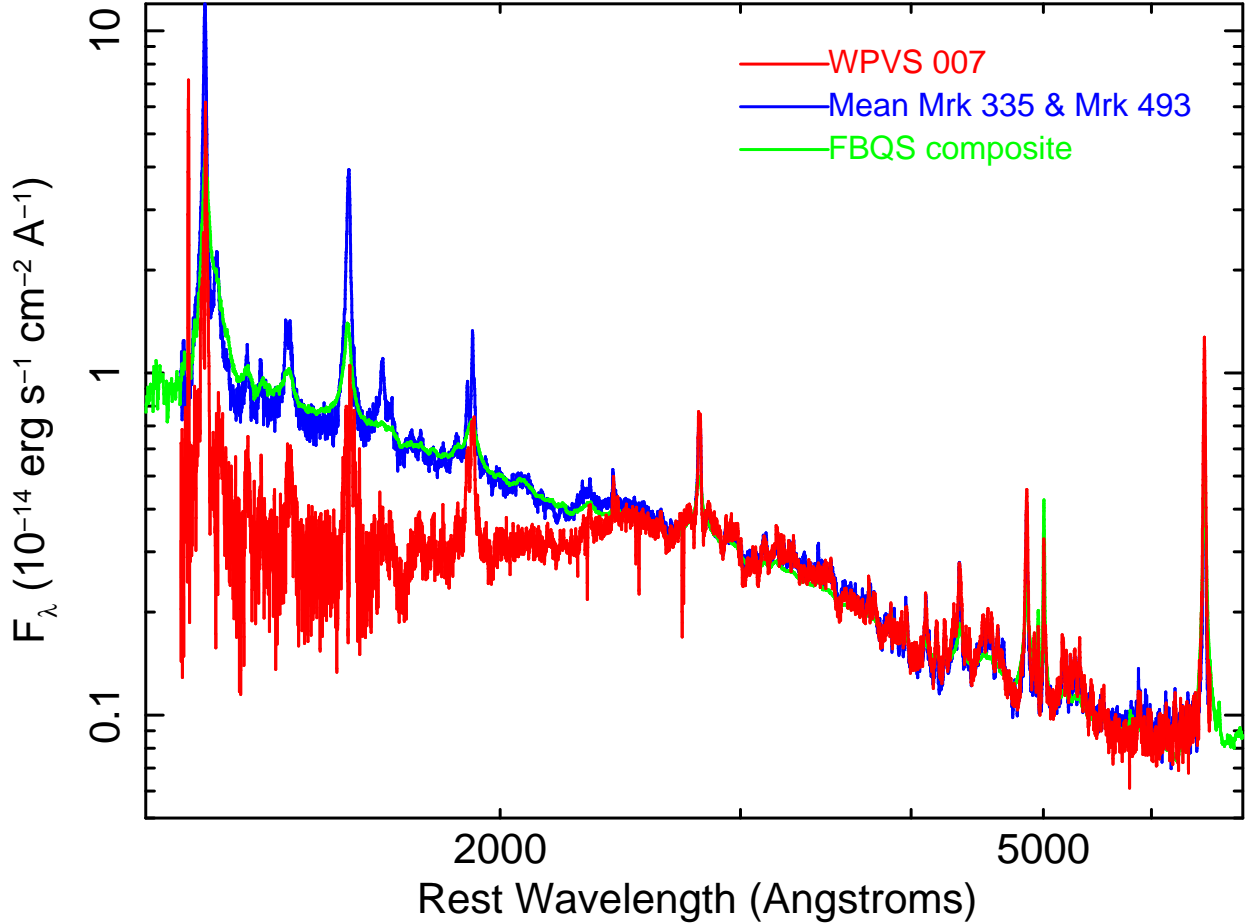}
\caption{The {\it HST} WPVS~007 spectrum (red) overlaid on the mean of
the {\it HST} spectra from two NLS1s Mrk~335 and Mrk~493 (blue) and
the FBQS radio-quiet composite spectrum 
\citep{brotherton01}.
\label{fig1}}
\end{figure}

The {\it HST} spectrum is relatively blue longward of $\sim
2700$\AA\/, but becomes significantly redder shortward of that value.
We show the spectrum in Fig.~\ref{fig1} overlaid on two comparison
spectra.  The first comparison spectrum is a mean of {\it HST} spectra
of two NLS1s that were chosen as follows with the intention of finding
objects similar to WPVS~007 in properties other than the absorption
lines and continuum shape.   As discussed in \citet{lm04}, 
there is significant range of UV emission-line shapes among NLS1s;
focusing on \ion{C}{4}, we found that some are rather narrow, strong
and symmetric around the rest wavelength, while others are broad,
weak and strongly blueshifted.  The miniBALs absorb most of the far UV
high-ionization emission lines in WPVS~007, but the emission lines
profiles outside of the absorption lines appear to be rather symmetric
about the rest wavelength.   Thus, we sought other NLS1s that had
emission lines similar in shape to the portion of the emission
lines in WPVS~007 outside of the absorption lines.    We found that
the WVPS~007 spectrum most closely resembled that of Mrk~493, which
had {\it HST} FOS spectra stretching from $\sim 1160$\AA\/ to
6818\AA\/ (observed frame), and Mrk~335, which had {\it HST} FOS
spectra stretching from $\sim 1160$\AA\/ to 3300\AA\/ (observed frame)
in this far UV region of the spectrum.  These two spectra were very
similar to one another in slope and emission lines in the overlapping
region.  These spectra were processed in the same way as discussed
above, and then uniformly resampled on a logarithmically-binned
wavelength scale that matched the original binning approximately.
They were then rescaled to match in flux and averaged over the
common-wavelength region.  The second comparison spectrum is the FBQS
radio-quiet composite spectrum \citep{brotherton01}.  The composite
spectrum has a very similar continuum shape as the Mrk~493--Mrk~335
average, implying that the Mrk~493--Mrk~335 have continua similar to
the average quasar. Thus, we make the simplifying assumption that the
unreddened intrinsic WPVS~007 spectrum has the same shape as the
Mrk~493--Mrk~335 average, and therefore similar to that of the average
quasar. 

\begin{figure}
\figurenum{2}
\epsscale{1.0}
\plotone{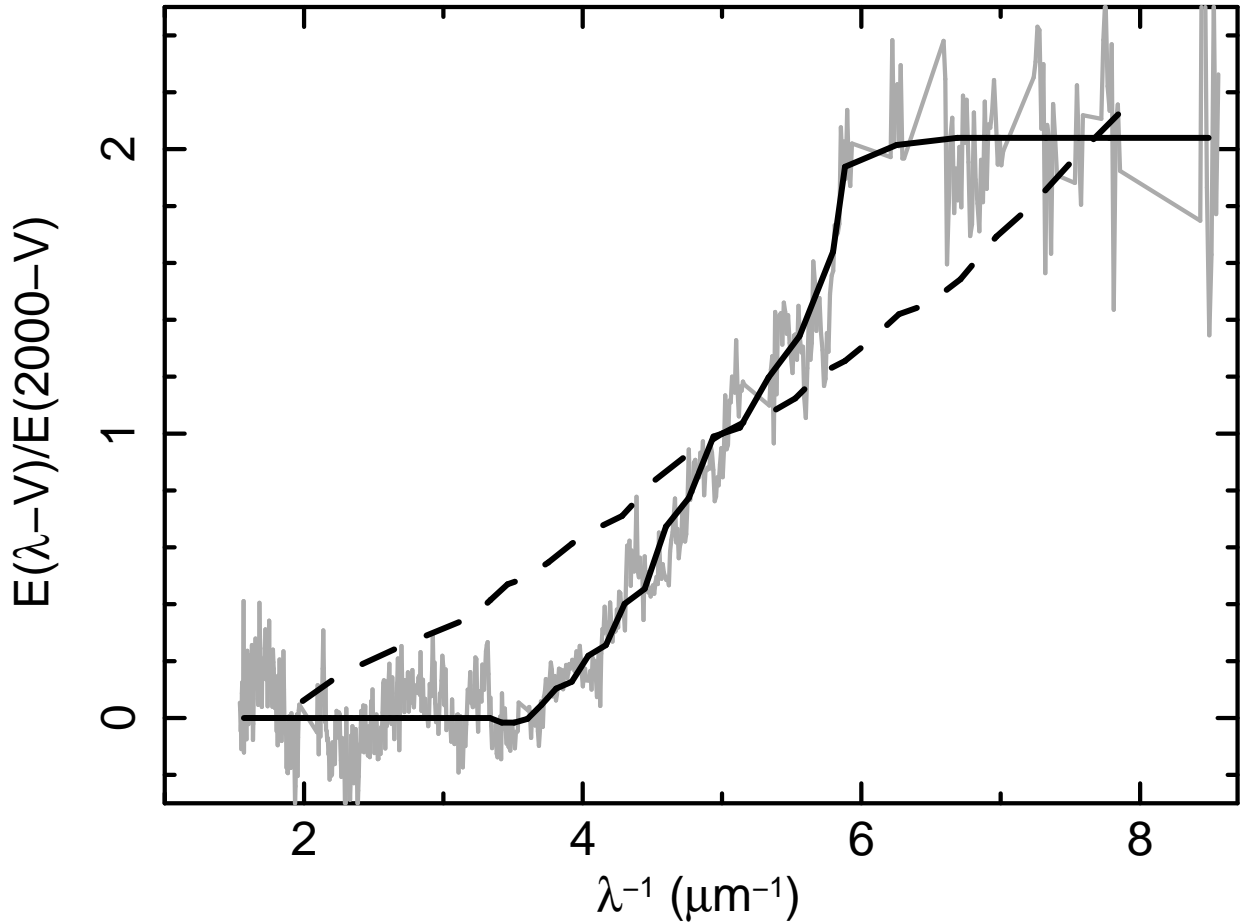}
\caption{Reddening curve for WPVS~007.  The grey line shows the ratio
  of the WPVS~007 spectrum with the average NLS1 spectrum obtained
  from Mrk~335 and Mrk~493.  The solid black line shows a spline fit
  to the ratio   spectrum.  The dashed black line shows the SMC
  reddening curve   \citep{prevot84}.\label{fig2}}
\end{figure}

Fig.~\ref{fig1} shows clearly that WPVS~007 is reddened in comparison
with the NLS1 mean spectrum and the composite spectrum.  The reddening
curve appears to be very unusual, however.  WPVS~007 and the
comparison spectra have essentially the same spectra longward of $\sim
2700$\AA\/, which means that there is either no attenuation in the
optical bands, or that there is no wavelength dependence in the
attenuation. We develop the 
reddening curve as follows.  We remove Galactic absorption lines and
then coarsely rebin the WPVS~007 spectrum and the Mrk~493/Mrk~335
average. We remove regions of prominent emission lines and compute the
ratio. The ratio was then fit with a spline model.  The ratio appears
to be approximately flat shortward of $\sim 1500$\AA\/.  The ratio
appears to have some structure longward of $\sim 3000$\AA\/, but it is
clear that is caused by higher equivalent-width optical \ion{Fe}{2} in
WPVS~007, so it was assumed to be 1 longward of $\sim 2700$\AA\/.
Since the ratio is flat in the optical band, we cannot define a
reddening curve in the standard way, relative to $E(B-V)$, as has been
done for other AGN by e.g., \citet{crenshaw01} and
\citet{crenshaw02}. Instead, we define it relative to 2000\AA\/, and 
assume that the optical bands are not attenuated at all.  The
extinction curve from the spectra, the spline fit and the SMC
reddening curve \citep{prevot84} are shown in Fig.~\ref{fig2}.  The
extinction curve in WPVS~007 is unusually steep between 2700\AA\/ and
1700\AA\/.  An attempt was made to reproduce this reddening curve
using silicate dust and a distribution of sizes \citep{wd2001}, but it
was unsuccessful (J.\ Weingartner, P.\ comm.\ 2008).  While the
unusual reddening curve is interesting, we note that it has no effect
on the absorption line measurements or analysis.

\subsection{Absorption Lines and Profile Analysis}

\begin{figure}
\figurenum{3}
\epsscale{1.0}
\plotone{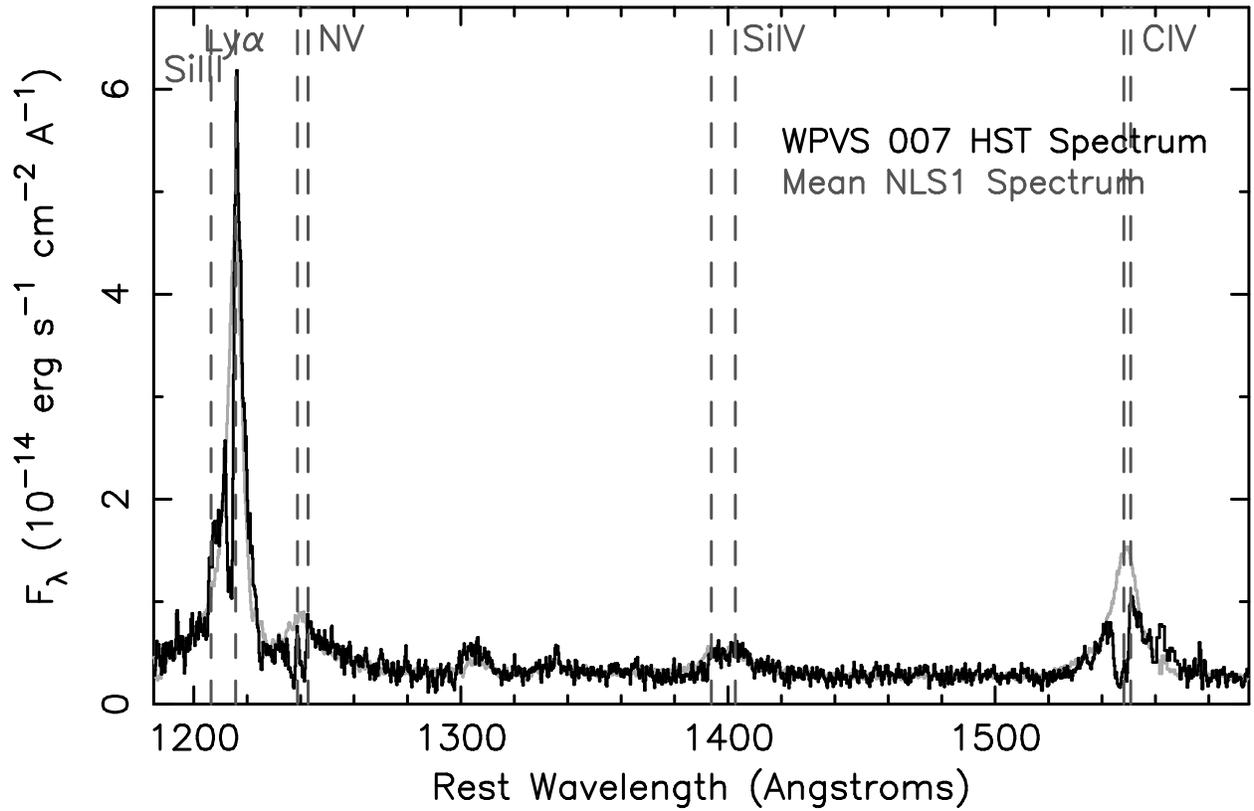}
\caption{The merged {\it HST} spectrum from WPVS~007 in comparison
  with the mean NLS1 spectrum created from the {\it HST} spectra of
  the NLS1s   Mrk~335 and Mrk~493.  Absorption lines are clearly present on
  Ly$\alpha$, \ion{N}{5}, and \ion{C}{4}.\label{fig3}}
\end{figure}

The {\it HST} spectrum reveals high-ionization absorption lines in
\ion{N}{5}, \ion{C}{4}, and Ly$\alpha$. Fig.~\ref{fig3} shows the
observed WVPS~007 spectrum (corrected for redshift and Galactic
reddening only) between 1185 and 1595\AA\/ in   comparison with the
average Mrk~335--Mrk~493 spectrum.    The presence of these absorption
lines was previously noted by \citet{crenshaw99}.    
The \ion{N}{5} absorption lines are clearly resolved, and we begin our
analysis there.  Preliminary examination of the lines shows that the
absorbing gas must occult both the line-emitting gas and the
continuum-emission region since the absorption lines are deeper than
the continuum level.   

We use a template analysis approach to analyze the absorption lines.
Briefly, we create an apparent optical depth template from the
\ion{N}{5} profiles and then fit that to the \ion{C}{4} and
Ly$\alpha$.  The template approach has strengths and weaknesses.  A
strength is that it is a very straightforward approach.  A weakness
is that, because of saturation and partial covering, it yields only a
lower limit on the column density of the absorbing ions.
Nevertheless, the results provide useful constraints on the properties
of the absorbing gas.

\begin{figure}
\figurenum{4}
\epsscale{1.0}
\plotone{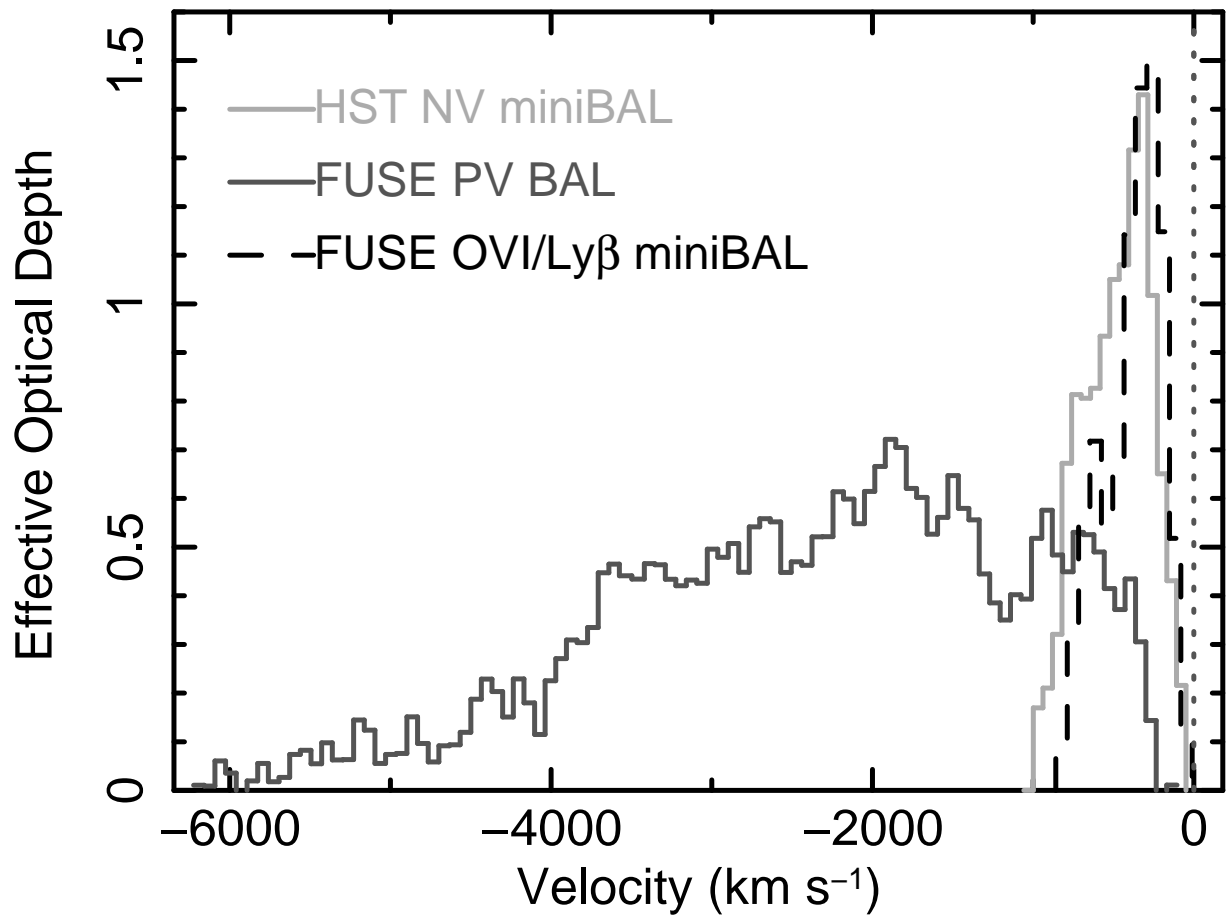}
\caption{The extracted optical depths as a function of velocity for
  the \ion{N}{5} mini-BAL, extracted from the {\it HST} spectrum (\S
  2.3),  the \ion{P}{5} BAL, extracted from the {\it FUSE} spectrum (\S
  3.1) and the \ion{O}{6}/Ly$\beta$ mini-BAL, extracted from the {\it
  FUSE} spectrum (\S 3.2).  Note the difference in onset velocity
  between the   BAL and the mini-BALs. There is some indication that
  the shape of the   mini-BAL changed between the {\it HST} and {\it
  FUSE} observations;   however, given the uncertainty in continuum
  placement and background   subtraction, those differences may not be
  significant.   \label{fig4}}
\end{figure}

We develop the apparent optical depth template as follows.  The mean
Mrk~335--Mrk~493 spectrum was resampled on the wavelength scale of the
WPVS~007 spectrum and the ratio $R$ of the WPVS~007 spectrum and the 
average spectrum was made.  Assuming that the average Mrk~335--Mrk~493
spectrum is a good representation of the unabsorbed WPVS~007 spectrum,
$R$ should be 1 outside of the  range of the absorption lines, and
should be less than one in the region of the absorption lines.  The
region of the \ion{N}{5} absorption lines was isolated, and the
apparent optical depths ($\tau = - \ln R$) for the two \ion{N}{5}
lines were computed.  The apparent optical depth profiles for the two
lines were essentially indistinguishable implying that absorption is
saturated but not black.  The shape is commonly understood to be the
consequence of a velocity-dependent covering fraction.  Since the
apparent optical  depth profiles were indistinguishable, they could be
averaged to produce a mini-BAL template (Fig.~\ref{fig4}).  The {\it
  HST} mini-BAL has an approximate maximum velocity of $V_{max} \sim
900 \rm \, km\, s^{-1}$ and approximate FWHM of $\sim 550 \rm \, km\,
s^{-1}$. 

\begin{figure}
\figurenum{5}
\epsscale{0.5}
\plotone{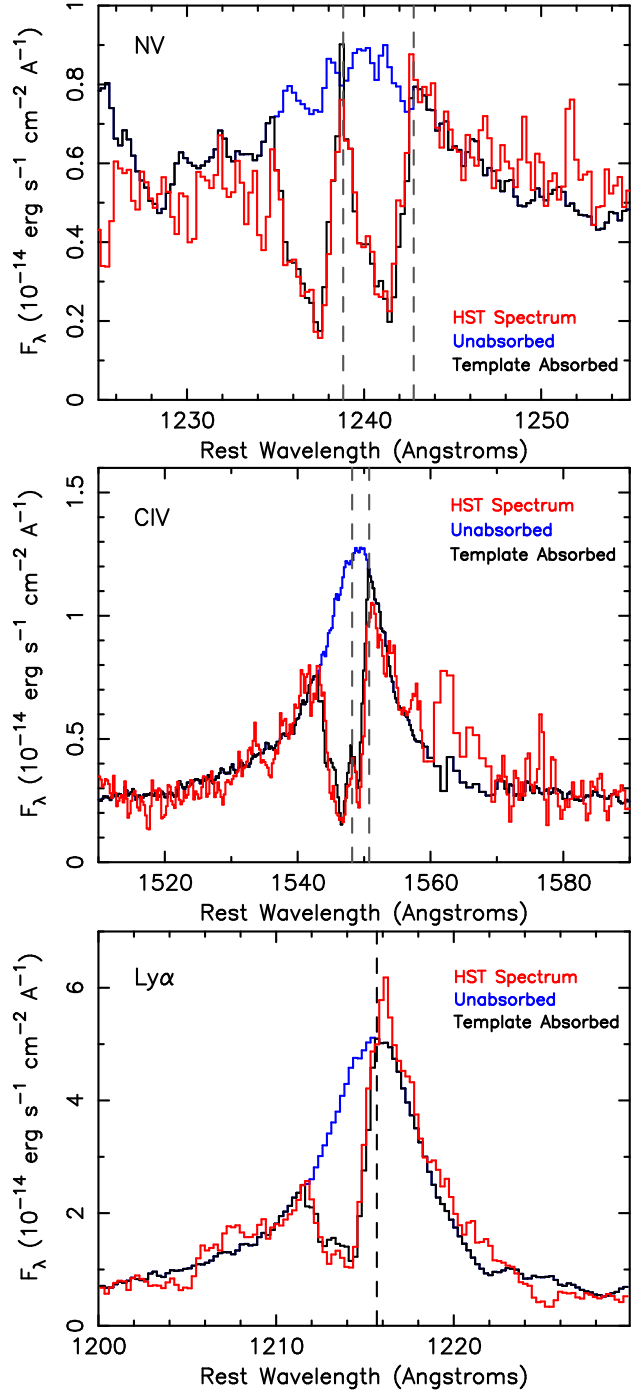}
\caption{Absorption lines in the {\it HST} spectrum (red lines)
  compared with the mean NLS1 spectrum (blue lines).  Top: The
  absorption line template was created   from the well-resolved
  \ion{N}{5} absorption lines using the   procedure described in \S 2.3.
  It was then applied to the mean NLS1   spectrum created from the the
  {\it HST} spectra of Mrk~335 and   Mrk~493 (black lines).   Assuming
  that the absorption lines are optically thick and have the same
  profile for all lines, we apply the template created from the
  \ion{N}{5} absorption lines to the \ion{C}{4} line (middle) and
  Ly$\alpha$ line (bottom). \label{fig5}}
\end{figure}

If the Mrk~335--Mrk~493 mean spectrum is an accurate representation of
the unabsorbed WVPS~007 spectrum, we should be able to apply the
mini-BAL template to the Mrk~335--Mrk~493 mean spectrum in the region
of \ion{C}{4} and Ly$\alpha$, and reproduce the observed WPVS~007
spectrum in those regions. To investigate this, we fit the spectrum
using IRAF {\tt   Specfit}.  {\tt Specfit}  allows a template
continuum, and for that we use the mean Mrk~493--Mrk~335 spectrum.
For the absorption lines, we use the template developed from  the
\ion{N}{5} line and shown in Fig.~\ref{fig4}.   We fix the template
wavelengths (in {\tt Specfit} treated as a redshift) to the rest
wavelengths for each line, and allow the scale factors to be free in
the spectra fitting.  The best fitting model results are shown in
Fig.~\ref{fig5} where we see that the \ion{N}{5} template reproduces
the \ion{C}{4} and Ly$\alpha$ absorption lines well.  The scale
factors are given in Table 2, which includes also the {\it FUSE}
spectral fitting  results discussed in \S 3.3.  The first column of
Table 2 lists the line, the  second column gives the vacuum
wavelength, the third column is the scale factor of the template
including the uncertainty from the spectral fitting, the fourth column
is the template used with the exception of the alternative deblending
of the {\it FUSE} spectrum as discussed in \S 3.3.4, and the fifth
column is the estimate of the lower limit of the column density
obtained by integrating over the line profile
\citep[Eq.\   9,][]{ss91}.  The uncertainties in the column densities
are proportional to the uncertainties in the scale factors in the
fitting, and thus they give only an estimate of the statistical
uncertainty.  We use the apparent optical depth to estimate
the column densities without accounting for saturation and partial
covering.  Therefore, the column densities are lower limits.
Atomic data were taken preferentially from
NIST\footnote{http://www.physics.nist.gov/PhysRefData/ASD/index.html},  
as well as from The Atomic Line List
v2.04\footnote{http://www.pa.uky.edu/~peter/atomic} and
\citet{morton91}.   

\section{{\it FUSE} Observation}

The observing log for the {\it FUSE} observation is given in Table 1.
The exposure times are split into ``day'' and ``night''
exposures. During the day exposures, the satellite is over the portion
of the earth in which the earth is lit by the sun.  Background from
scattered light is somewhat higher during the day exposures, and
emission from airglow is stronger.  The LIF detectors have larger
effective area, and we used both the day and night data from them. The
SIC detectors have significantly lower effective area; thus, since
WPVS~007 is a somewhat faint AGN, we used only night data from them.
Hence, the exposure times reported in Table 1 are shorter for the SIC
spectra compared with the LIF spectra.

The {\it FUSE} data were processed with a modified version of the
CalFUSE pipeline version 3.1.8.  The modifications involved a
reduction of background intensity using a PHA selection. The details
are given in Appendix A.  The resulting spectra were then merged and
smoothed, and the details of that process are given in Appendix B.
The resulting spectrum is shown in the top panel of
Fig.~\ref{fig6}. Identified Galactic absorption lines are labeled below
the plot, and the rest wavelengths of common absorption lines are
labeled above the plot, regardless of whether they are detected 
or not. A plausible unabsorbed continuum using the {\it HST} quasar
composite spectrum \citep{zheng97} is included on the graph.

The spectrum clearly shows the presence of broad absorption lines in
both high-ionization lines such as \ion{O}{6}$\, \lambda\lambda 1031.9,
1037.6$ and low-ionization lines such as \ion{C}{3}$\,\lambda 977$.
There are a few features that are seen immediately, and they are 
confirmed by the detailed analysis below.  \ion{P}{5}$\, \lambda\lambda
1118, 1128$ is present. This line from the rather low-abundance
element phosphorus has been shown to be characteristic of
absorption-line gas with high column density \citep{hamann98}.  The  
1118\AA\/ component is clearly deeper than the 1128\AA\/ component,
suggesting that the lines are perhaps not completely saturated and
thus the optical depth is not extremely high, and that partial
covering is present. \ion{O}{6} absorption is
dominated by a broad and deep trough, but narrow absorption lines
similar to the mini-BALs seen in the {\it HST} spectra are observed in
both the \ion{O}{6} doublets and also in Ly$\beta$.  The mini-BALs are
not apparent in the \ion{P}{5} troughs.   

Before we perform a quantitative analysis of the absorption lines in
WPVS~007, we qualitatively compare our spectrum with the {\it HST}
spectrum of a BALQSO LBQS~1212+1445, also shown in the lower panel of
Fig.~\ref{fig6}.  At a redshift of $z=1.6245$, the {\it HST} STIS
G230L bandpass samples the restframe 600--1200\AA\/ band. This
comparison is useful because the 10,811-second-exposure {\it  HST}
spectrum has considerably better signal-to-noise ratio than our  {\it
 FUSE} spectrum and thus the lines are easier to identify.    In
LBQS~1212+1445 we see many of the same lines as in WPVS~007. The
\ion{P}{5} BALs are clearly present, and the 1118\AA\/ 
component is  clearly deeper than the 1128\AA\/ suggesting again that
the line is not very highly optically thick and that partial covering
is important. The velocity profile has a larger $v_{min}$ but perhaps
the same $v_{max}$ for \ion{P}{5}.  Minibals are present in \ion{O}{6}
in both objects, and are also clearly present in \ion{C}{3} and
\ion{N}{3} in LBQS~1212+1445. 

\subsection{\ion{P}{5} Analysis}

Analysis of broad absorption lines is very complicated.   If
the lines are overlapping, the continuum may be difficult to
identify, and  absorption lines themselves may be blended.  The
opacity of a line depends on velocity, there may be partial
covering that may also depend on velocity, and these cannot be
diagnosed directly because of blending.  The covering fraction can
also differ for different lines and ions.   In the case of very high
signal-to-noise spectra and unblended lines, these factors may be
robustly deconvolved \citep[e.g.,][]{arav07}, but that is not possible
for our relatively blended, low-signal-to-noise spectrum.  Therefore,
our goal in this section is to use the \ion{P}{5} $\lambda \lambda
1117.98, 1128.01$ feature in the {\it FUSE} spectrum to derive an
apparent optical depth profile to be used to fit and deblend the other
lines.  

As noted in \S 3.1, \ion{P}{5} is clearly seen in the {\it FUSE}
spectrum.  Phosphorus has much lower solar abundance than other
elements such as carbon, nitrogen and oxygen that commonly produce
BALs \citep[e.g., $\rm P/C=0.00093$,][]{grevesse07}.  Therefore, even
if those lines are saturated and partial covering is important, it is
possible that the \ion{P}{5} will not be as optically 
thick.     \ion{P}{5} occurs in a region of the spectrum where
absorption from other ions is not generally observed, and therefore it
may not be subject to severe blending. In fact, the doublet ratio
(1118\AA\/ component divided by the 1128 \AA\/ component) is 2.03, and
visual examination of the region suggests that the 1118\AA\/ component
has a  larger apparent optical depth than the 1128\AA\/ line. 
Therefore we first analyze the \ion{P}{5} line with the intention of
developing a apparent absorption line template from this feature that
can be used for the other absorption lines.

\begin{figure}
\figurenum{6}
\epsscale{1.0}
\plotone{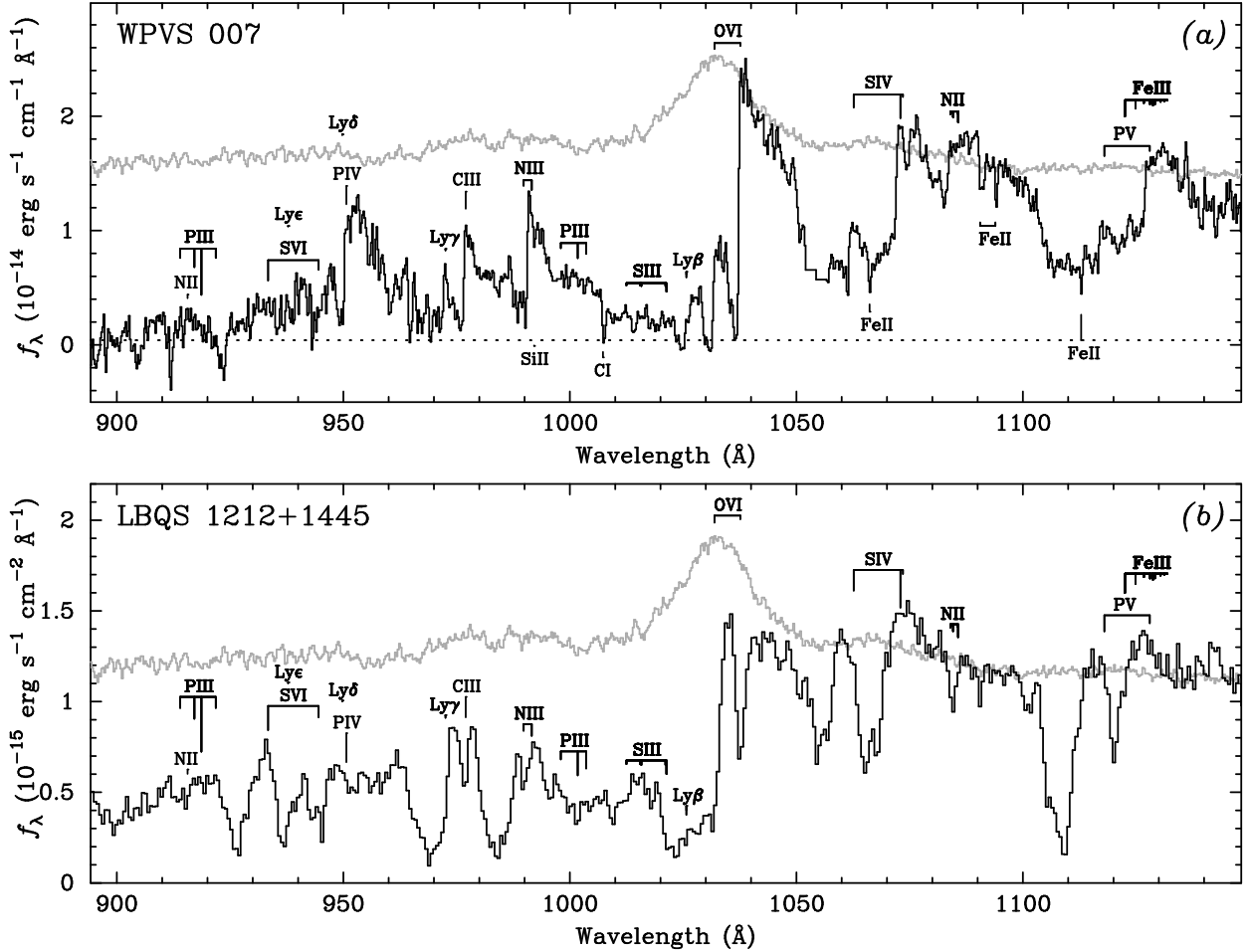}
\caption{(a) The merged, smoothed, dereddened, and deredshifted {\it FUSE}
  spectrum of WPVS~007 (black line).  For comparison, the scaled {\it
  HST} quasar composite spectrum  is shown.  Identified Galactic
  absorption lines are marked below the spectrum.  Rest wavelengths of
  prominent emission lines in   the bandpass are labeled above the
  spectrum, regardless of whether   or not the corresponding
  absorption is observed.  Length of tick marks are proportional to 
  $gf$ for individual multiplets. (b.) The {\it HST} STIS spectrum of
  the \ion{P}{5} BALQSO LBQS~1212$+$1445 for comparison.  \label{fig6}}
\end{figure}

\begin{figure}
\figurenum{7}
\epsscale{1.0}
\plotone{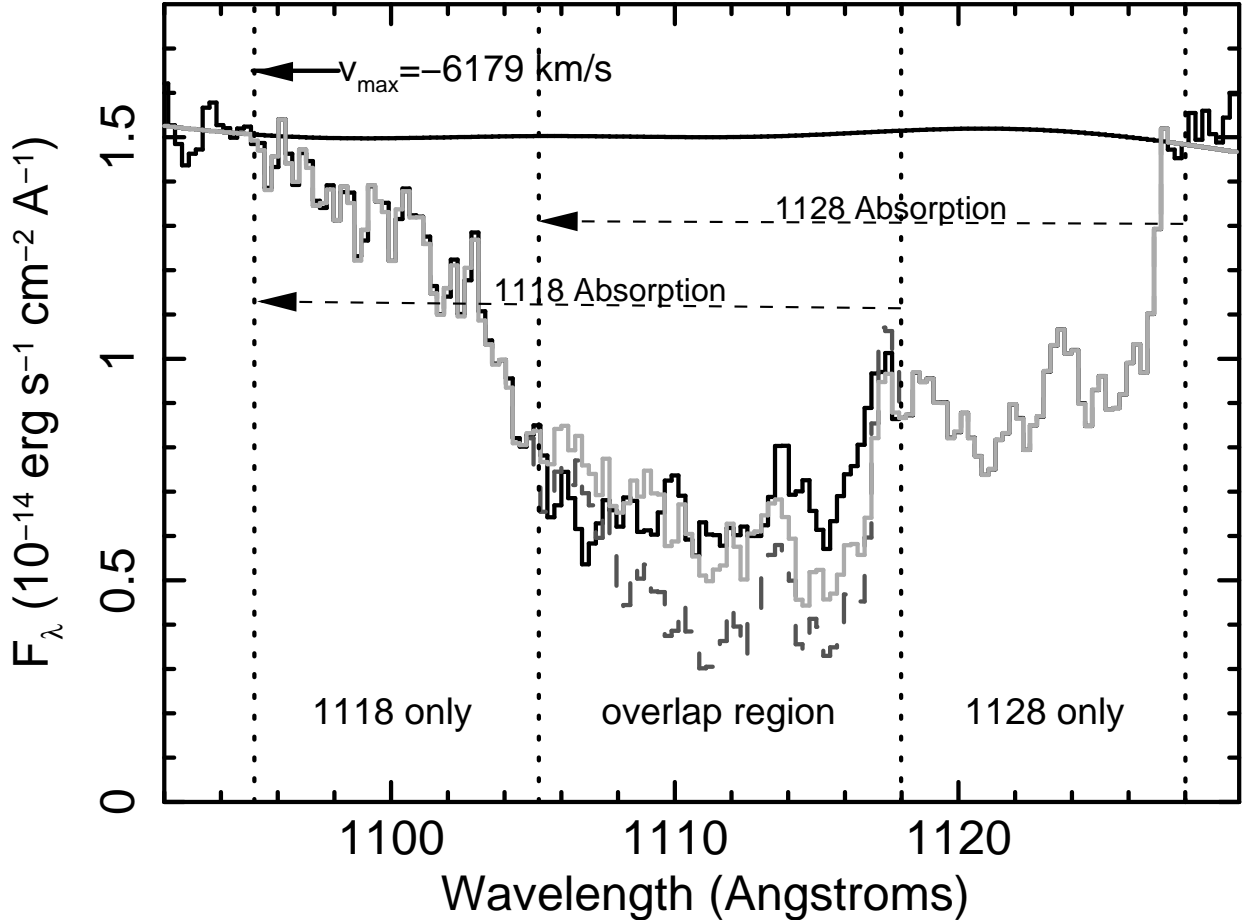}
\caption{The analysis of the \ion{P}{5} region of the {\it FUSE}
  spectrum.  The spectrum is shown by the black line. The continuum, a
  spline fit to the scaled {\it HST}   composite spectrum
  \citep{zheng97}, is shown by the nearly horizontal black   line.
  The vertical dotted lines delineate the regions of the absorption
  profile either  represented by only the 1118\AA\/ component of
  \ion{P}{5} (left), by   only the 1128\AA\/ component of \ion{P}{5}
  (right), and by both   (middle).  $V_{max}$ shows the high-velocity
  extent of the BAL inferred from the 1118\AA\/ component.  The dashed
  dark grey line shows   the inferred spectrum if the apparent optical
  depth of the 1118\AA\/   component is  2.03 times the apparent
  optical depth of the 1128\AA\/   component, as would be  appropriate
  if the gas were optically thin and partial covering were negligible.
  The light grey line shows the results of apparent optical depth
  fitting described in \S 3.1.  The ratio of the apparent optical
  depths was 1.35.  See text for details.\label{fig7}} 
\end{figure}

Fig.~\ref{fig7} shows an expanded version of the region of the {\it FUSE}
spectrum near \ion{P}{5}, after three  Galactic absorption lines
marked in Fig.~\ref{fig6} had been modeled out.   For the continuum,
we use a spline fit to the {\it HST} FOS composite spectrum with an
intial factor of 0.7 scaling (the level shown in Fig.~\ref{fig6}).  We
assume that that \ion{P}{5} is solely responsible for the absorption
between 1095 and 1127\AA\/.  Therefore, the termination of the
absorption at 1095\AA\/ marks the maximum extent of the 1118\AA\/
absorption trough, implying a maximum velocity of $6179\rm \, km\,
s^{-1}$ for the broad absorption lines.  Using this fact, the region
between 1095 and 1127\AA\/ can be broken into three parts: one in
which the absorption is produced by the 1128\AA\/ component only
(right on Fig.~\ref{fig7}), by the 1118 \AA\/ component only (left on
Fig.~\ref{fig7}), and by a combination of both components (middle on
Fig.~\ref{fig7}).  The extents of the 1128\AA\/ absorption trough and
the 1118\AA\/ absorption trough are also marked on Fig.~\ref{fig7}.  

In terms of velocities, we glean the following information from 
Fig.~\ref{fig7}.  The velocity profile between zero and $-2678\rm \,
km\, s^{-1}$  can be constructed from the 1128\AA\/ component (right
region in Fig.~\ref{fig7}).  The velocity profile between 3555 and
$6179\,\rm km\,s^{-1}$ can be constructed from the 1118\AA\/
component (left region in Fig.~\ref{fig7}).  Apparent optical depth
profiles from these two regions are 
easily obtained. But the velocity profile between $-2678$ and
$-3555\rm \, km\, s^{-1}$ must be constructed from the overlap
region.  The overlap region consists of a superposition of absorption
from the  $-2678$-- $-3555\rm \, km\, s^{-1}$ part of the 1128\AA\/
trough and the 0-- $-854\rm \, km\, s^{-1}$ part of the 1118\AA\/
trough.  

Thus our strategy to produce the apparent optical depth profile is as
follows. We generate the apparent optical depth by dividing the
spectrum by the continuum and taking the logarithm ($\tau_{eff}=-\ln
R$).  We construct the profile in the 0--2678$\,\rm km\, s^{-1}$ band
using the 1128\AA\/ component, and the profile in the 3555--6178$\,\rm
km\, s^{-1}$ using the 1118\AA\/ component.  Then, the two estimates
for the 2678--3555$\,\rm km\, s^{-1}$ are generated from the overlap
regions by subtracting the known component from the apparent optical
depth.  The overlap region is then the mean of these two estimates.
To test the resulting template, we apply it to the continuum to
generate the spectrum.

The two  uncertainties in this procedure are the ratio of the apparent
taus of the two lines and the placement of the continuum.
Specifically, we assume that the derived optical depth profile
corresponds to the 1128\AA\ component, and the 1118\AA\/ component is
related by a ratio, to be determined. Initially, we assume that the
ratio is the ratio of the $gf_{ik}$ for the two lines, 2.03,
appropriate if the absorption lines were optically thin and the
covering fraction is unity.   The result is 
shown as the dashed dark grey line in Fig.~\ref{fig7}.  By definition,
the simulated spectrum models the profile well in the low and high
velocity regions of the feature that are created by either the 1118 or
the 1128\AA\/ component each by itself. However, in the overlap
region, the synthetic spectrum clearly shows stronger absorption than in the
observed spectrum.     

There are several possible origins for the discrepancy between the
synthetic spectrum and the observed spectrum.  First, the continuum
level could be lower than assigned. A continuum offset seems unlikely,
though, as we set the continuum level at the point where the sharp
low-velocity drop on the 1128\AA\/ component occurs.  More likely, the
absorption lines could be saturated ($\tau >1$)  and partial covering
could be important so that the ratio between the apparent optical
depths of the 1118 component and the 1128 component is less than 2.03
and approaches 1.  

Since the ratio of the apparent taus of 2.03 is ruled out, we
determine a best fitting ratio as follows.  We define a figure of
merit (fom), which is the sum of the absolute value of the difference
between the simulated spectrum and the observed spectrum in the
overlap region.  Simulated spectra are generated using a range of
apparent tau ratios  between 2.03 and 1.  The fom is a shallow
apparently parabolic function of the ratio with a minimum of 1.35. The
simulated spectrum using this best fitting apparent tau ratio is
shown by the light grey line in Fig.~\ref{fig7}.  The fit is not perfect,
being too shallow for shorter wavelengths and too deep for longer
wavelength in the overlap region, but it is clearly better than the
fit for a ratio of 2.03. Another possibility is that the apparent tau
ratio varies as a function of velocity, with a lower ratio
(approaching 1)  for low velocities and a higher ratio for high
velocities.  Since the fom has no statistical significance, we
estimate the uncertainty on this value by examining at the residuals
and determining the point at which the simulated spectrum clearly does
not fit the data, yielding a  conservative range of 1.15--1.55.     

The constraints on reasonable placement of the continuum are fairly
tight.  The absorption decreases sharply at low velocity, and thus the
continuum cannot reasonably be placed very far below the position
where this joins the continuum.  Likewise, unless we do not see the
continuum at all in this object, we cannot place the continuum much
higher.  We examine the cases in which the continuum was set to be
about 4\% higher and lower than the original value; this range seemed
consistent with the constraints above.  The results are very similar.
A higher continuum yields a higher maximum velocity and perhaps a
slightly better fit for longer wavelengths in the overlap region.  A
lower continuum yields a lower maximum velocity and a notably worse
fit at longer wavelengths in the overlap region.  The best {\it fom}
was obtained for the original continuum level, but the differences are
not large.  Therefore, we use the optical depth profile obtained using
the original continuum level, with an inferred ratio of the apparent
optical depths of the 1118\AA\/ component to the 1128\AA\/ component
of 1.35. The apparent optical depth profile plot is shown in
Fig.~\ref{fig4}. The profile has an approximate maximum velocity of
$V_{max} \sim 6000 \rm \, km\, s^{-1}$, and an approximate FWHM of
$\sim 3400\rm \, km\, s^{-1}$. 

As shown in Fig.~\ref{fig6}, a set of \ion{Fe}{3} transitions occurs
near 1126 \AA\/, between the two \ion{P}{5} lines.  We do not believe
that this transition produces significant absorption for several
reasons.  First, if there were significant absorption from
\ion{Fe}{3}, the template that we derive would be wrong; yet, as we
show below, this template produces an excellent fit for the \ion{S}{4}
absorption features at 1063 and 1073 \AA\/.  In addition, we noted the
similarity to the LBQS~1212$+$1445 spectrum also shown in
Fig.~\ref{fig6}.  The \ion{P}{5} troughs appear narrower and are not
blended with one another in that object, and \ion{Fe}{3} absorption
does not appear to be present.   For future use,  we measure upper
limits on the column density of a multiplet of \ion{N}{2} near
1085\AA\/, and one of \ion{Fe}{3} near 1126 (Fig.~\ref{fig6}) by
inserting a {\it FUSE} BAL  template into the model and increasing the
normalization until the $\chi^2$ increases by 6.63, corresponding to
99\% confidence for one parameter of interest. These upper limits are
also listed in Table 2.  

\subsection{MiniBAL Analysis}

\begin{figure}
\figurenum{8}
\epsscale{0.6}
\plotone{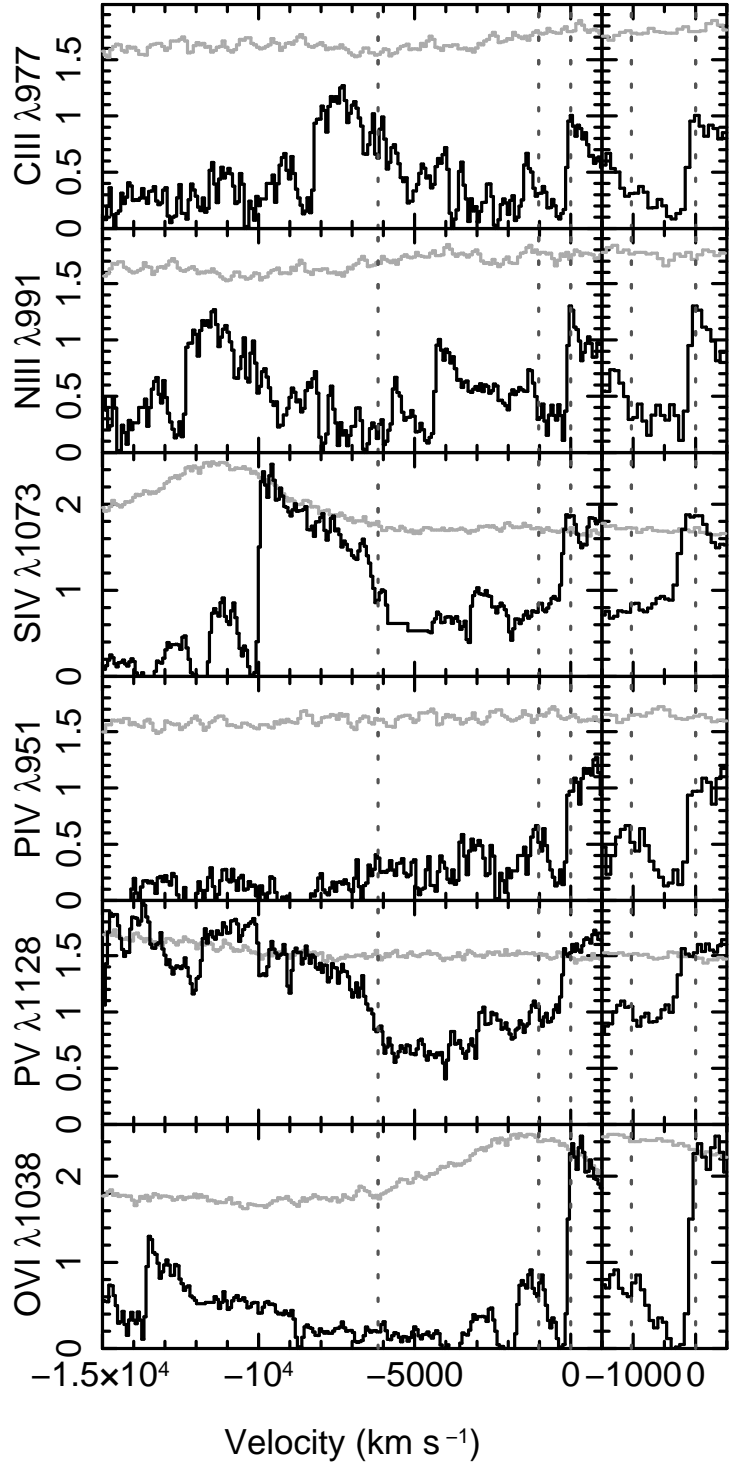}
\caption{The {\it FUSE} spectrum plotted as a function of velocity
  relative to various absorption line rest wavelengths.  The light
  grey lines show the scaled {\it HST}  composite spectrum
  \citep{zheng97}.  The right side shows an 
  expanded view of the low-velocity region.   Note the overall
  similarity between the \ion{S}{4} and the \ion{P}{5} profiles; the
  absorption blueward of \ion{O}{6} is much broader.  In addition, the
  onset of the \ion{P}{5} and \ion{S}{4} absorption occurs at higher
  velocity   than the other lines, indicating that the mini-BAL is not
  present in   \ion{P}{5} and \ion{S}{4}.\label{fig8}}
\end{figure}

The {\it FUSE} spectrum of WPVS~007 has  mini-BALs as well as
BALs. These can clearly be seen on \ion{O}{6} and Ly$\beta$, and they
can also be seen on \ion{N}{3} and \ion{C}{3}.  They are not seen on
\ion{Si}{4} or \ion{P}{5}.  They are distinguished from the
BALs by their distinctive shape, most clearly seen on \ion{O}{6} but
also because they have a lower-velocity onset than the BALs, as shown
in in Fig.~\ref{fig8}.  In this section we analyze the mini-BALs in the
region of Ly$\beta$ and \ion{O}{6}.   We note that although the
miniBALs are not blended, we are not able to derive the optical depth
and  covering fraction as a function of velocity, both because of the
poor signal-to-noise ratio and because of a zero-offset problem
(described below).  Thus, as before, our goal is to produce and
apparent tau velocity profile to be used as a template for fitting the
other miniBALs.  

The mini-BALs are superimposed on the BALs, and thus their profiles
are difficult to analyze.  To define a pseudocontinuum, we
isolate the region of the  spectrum between 1020 and 1038\AA\/, and
remove the regions of the spectrum containing the mini-BALs from
Ly$\beta$, and \ion{O}{6}$\,\lambda\lambda 1031.9, 1037.6$.  We then
fit the remainder with a model intended to parameterize the 
pseudocontinuum. We find that a lorentzian line  model plus a constant
works well.  Next, we need to find the apparent optical depths, generally
done by dividing the continuum by the data, and then taking the
logarithm.  The problem with that procedure for these data is that the
spectrum in the region of the mini-BALs dips below zero.  This is, of
course,  not physical, and may be a consequence of residual background
subtraction problems, although as shown in figures in Appendix A, the
errors are sufficiently large that the data are consistent with zero
in the region of the mini-BALs. Therefore, to obtain the template, we
add 0.1 to both the continuum model and the spectrum before taking the
logarithm.  The result is encouraging, as it shows three apparent
optical depth profiles that are very similar in shape and depth,
although the \ion{O}{6}$\,\lambda 1032$ component has slightly larger 
apparent optical depth.  We resample the Ly$\beta$ and \ion{O}{6}$\,\lambda
1038$ components to the wavelengths of the \ion{O}{6}$\, \lambda 1032$
component, and compute the mean profile.  The result, as a function of
velocity, is shown in Fig.~\ref{fig4} in comparison with the mini-BAL template
developed from \ion{N}{5} in the {\it   HST} spectra, and the
{\it FUSE} BAL profile developed in \S 3.3. {\it FUSE} mini-BAL
template has  approximately  the same maximum apparent optical depth
as the {\it HST} mini-BAL; however, this is a lower limit as we had to
add a constant to the spectrum in order to compute the apparent
optical depth.  The {\it FUSE} mini-BAL seems to perhaps have a
sharper onset, a little lower maximum velocity ($\sim 800 \rm \,
km\, s^{-1}$) and a little smaller FWHM ($\sim 470 \rm \, km\,
s^{-1}$), compared with the {\it   HST} mini-BAL, perhaps indicating
evolution of this component.   

\subsection{Modeling the FUSE Spectrum}

Now that we have developed apparent optical depth templates for the
BAL (discussed in \S 3.1) and the mini-BAL (discussed in \S 3.2), we
can apply the templates to the remainder of the {\it FUSE} spectrum.
First, though, we emphasize that the template approach has both values
and shortcomings.  The templates can be used to deblend the complicated
absorbed spectrum to identify which lines are present and obtain
estimates of the absorption columns of associated ions.  But since we
have only the apparent optical depth profile, and no information about
the covering fraction that may be both velocity and ion dependent, we
can only obtain lower limits on the ionic column densities.  In
addition, we make the assumption that all of the lines have the same
apparent optical depth profile, while in reality, the higher
ionization lines are sometimes observed to have broader profiles in
other objects.  We address this final point in \S 3.3.4.

Our approach is as follows.  As for the {\it HST} spectrum, we fit the
spectrum using IRAF {\tt Specfit}.  We use  the {\it HST} composite
continuum \citep{zheng97} with the normalization fixed at the level
shown in Fig.~{6} for the continuum.  We use the two templates derived
in \S 3.1 and 3.2, fixing the wavelength (treated in {\tt Specfit} as
a redshift) to correspond to the rest wavelengths of the absorption
lines investigated, allowing only  the scale factor (effectively,  the
apparent optical depth) to be free. To obtain the best fit, we allow
the apparent optical depth ratios within multiplets to deviate from
the values prescribed by atomic physics.  This leads to separate
limits on the ionic column densities derived from each line.   We
discuss three regions of the spectrum separately: the
\ion{P}{5}/\ion{S}{4} region between 1037.5 and 1135\AA\/, the
\ion{O}{6} region between 991 and 1135\AA\/, and the
\ion{C}{3}/\ion{N}{3}/\ion{P}{4} region shortward of 991\AA\/.    

\subsubsection{\ion{P}{5}/\ion{S}{4} Region}

We first fit the \ion{P}{5} region by itself, between 1090 and
1135\AA\/.  The results for the fit of the entire {\it FUSE} spectra
are given in Table 2, where the definitions of the columns were
discussed in \S 2.3. The best fit is 
shown in Fig.~\ref{fig9}.  The upper panel shows the final model
superimposed upon the spectrum, and the lower panel shows each
absorption component.

We next fit the region between 1037.5 and 1135 \AA\/. This region
includes a large absorption feature that can plausibly be identified
as \ion{S}{4}.  This line is composed of three components.  The ground
state transition has a wavelength of 1062.7\AA\/. The other two
transitions from this configuration arise from an excited state with
$E_i=951.43 \rm \, cm^{-1}$, with wavelengths 1072.96 and
1073.51\AA\/.    The $g_i f_{ik}$ values for the  three transitions
are 0.69, 2.09, and 0.23.  The latter two lines (the transitions from
the excited state) are very close together in wavelength, and we model
them with the $g_i f_{ik}$-weighted mean wavelength of
1073.07. Absorption from ions in the excited state is clearly present;
it can be recognized by the sharp low-velocity onset near 1073\AA\/
(Fig.~\ref{fig6}).  We obtain an excellent fit, as seen in
Fig.~\ref{fig6}.  In particular, the position of the sharp increase in
opacity at low velocity matches the spectrum perfectly for both
components, verifying that this feature is indeed \ion{S}{4} and that
significant absorption from the excited state is present

The undeniable presence of the excited state absorption from
\ion{S}{4} is interesting because the energy of the lower level is
slightly (0.118 eV) compared with typical values of nebular $kT$.
Population of this level requires $n_e > A_{ki}/q_{ki}$, where
$A_{ki} = 7.70\times 10^{-3}$ is the Einstein A value for 
the transition from the excited state to the ground state, resulting
in the \ion{S}{4} $10.51\, \mu m$ line, and $q_{21}$ is the collision
deexcitation rate, which can be found from the collision strength
$\Upsilon$ for this transition. For a nebular temperature of 15,000 K,
\citet{ss99} give $\Upsilon($\ion{S}{4})$=8.48$.  The resulting
critical density is $n_{crit}=4.7\times 10^4\,\rm cm^{3}$.  The
fitting reveals that the excited-state line is somewhat stronger than
the ground-state line (scale factor $1.54 \pm 0.03$ vs $1.20 \pm
0.05$), yielding a ratio of 1.28.  At densities above critical, the ratio
would approach $g_{ik}(excited)/g_{ik}(ground)=2.02$.  This suggests
that the gas density is near or above the critical density.

\subsubsection{The \ion{O}{6} Region}

Between 991 and 1037.5\AA\/ there is a very broad absorption feature
as well as the mini-BALs originating in \ion{O}{6} and Ly$\beta$ that
were discussed in \S 3.2.  \ion{O}{6} is at least partially
responsible for the broad feature.  However, if the \ion{O}{6}
components have the same profile as the \ion{P}{5} components, and if
\ion{O}{6} were the only contribution to this feature, then the
high-velocity decrease in opacity should be observed at approximately
1016\AA\/ (the point corresponding to 20\% of the maximum opacity of
the profile shown in Fig.~\ref{fig6} or \ref{fig9}), rather than
stretching to at least 991\AA\/ as observed.  This may be interpreted
as evidence that  there are several other ions responsible for
absorbing the shorter wavelengths.  In this section, we investigate
this possibility by including absorption from ions that are
conceivably present in the gas.  On the other hand, the \ion{O}{6}
profile could be alternatively interpreted as being  broader than the
\ion{P}{5} profile; this would not be surprising if true  as the
profiles of high-ionization lines are frequently found to be broader
than those of intermediate ionization lines \citep{jchs01}. Therefore,
we discuss an alternative deblending in \S 3.3.4. 

In this section, we take the approach that the \ion{O}{6} BAL
resembles the \ion{P}{5} BAL.  It is clear from the discussion
above that we will not be able to model the entire feature using the
\ion{O}{6} BAL alone, but before adding other lines, we first determine the
shortest wavelength within the feature that we can obtain an adequate
fit assuming the presence of only an \ion{O}{6} BAL and miniBAL.  We
find that we can obtain an adequate fit down to 1018\AA\/, although
the fit is not very good between the Ly$\beta$ and the
\ion{O}{6}~$\lambda 1032$ mini-BAL because the \ion{O}{6} BAL is
required to be very deep ($\tau=4.3$) to model all of the deep
absorption present.  Decreasing the lower limit of the fitting range
further yields larger $\chi^2$ and larger $\tau$ for the
\ion{O}{6}~$\lambda 1032$  BAL. 

We observe a Ly$\beta$ mini-BAL, so it is quite possible that a
Ly$\beta$ BAL is also present.   We add this component, and find a
much improved fit around the mini-BALs.  We find that the fit is
adequate down to 1012\AA\/; for wavelengths shorter than that, the
$\chi^2$ rises precipitously, and the broad Ly$\beta$ optical depth
becomes very large.

\ion{S}{3} has resonance and low-lying excited state transitions in
this region of the spectrum.  The multiplet consists of six
components at 1012.50, 1015.50, 1015.57, 1015.78, 1021.11, and
1021.32\AA\/, with corresponding lower level energies of 0, 298.7,
298.7, 833.1, and 883.1$\, \rm cm^{-1}$, and $g_i f_{ik}$ values of 0.042,
0.042, 0.032, 0.053, 0.053, and 0.158, respectively.  We model this
feature with three BALs centered at 1012.50, 1015.63, and
1021.27\AA\/.  We obtain a good fit for wavelengths down to about
1000\AA\/. We note that
as we add components to the model, the uncertainties on the optical
depths increases; for example, the uncertainty on the \ion{S}{3}$\,
\lambda 1015$ component is almost as large as the optical depth.  When
the uncertainty is as large as the measurement, the detection is less
than robust.  Yet, if we maintain the assumption that the entire broad
feature between 991 and 1037.5\AA\/ is comprised of features with the
same profile as \ion{P}{5}, the \ion{S}{3} component is necessary.  

The \ion{S}{3} can account for the deepest part of the trough for
wavelengths as short as 1000\AA\/; however, shortward of that
wavelength, its opacity drops rapidly. If the broad feature between
991 and 1037.5\AA\/ is a blend of lines with the same profile as
\ion{P}{5}, another ion with resonance wavelength around 1000\AA\/
must participate in the absorption.  We found a \ion{P}{3} resonance
triplet  that will work.  The wavelengths are 998.0, 1001.726, and
1003.6\AA\/, with corresponding $gf$ values of 0.223, 0.668, and
0.473. Including these allows a good fit to the shortest wavelength of
this feature at 991\AA\/.  The results of this fit are given in Table
2.  However, the shortest wavelength component
ends up having zero normalization.  We obtain an upper limit by
increasing the normalization of this component until
$\Delta\chi^2=6.63$, corresponding to 99\% confidence for one
parameter of interest. The derived column density upper limit for
\ion{P}{3} $\lambda 998.0$ is comparable to the lower limits for the
other two components. 

\subsubsection{The \ion{C}{3}, \ion{N}{3} and \ion{P}{4} Region}

We finally turn to the region of the spectrum shortward of
991\AA\/ that contains \ion{C}{3}, \ion{N}{3} and \ion{P}{4}.
Starting with the region between 955 and 991\AA\/, we see clear
evidence for both broad components and mini-BALs for \ion{C}{3} and
\ion{N}{3}, so we begin by adding those lines.  The region around
\ion{N}{3} was fit very well with a BAL and a mini-BAL.  In the
\ion{C}{3} region, the mini-BAL template modeled the region between 972
and 977 angstroms well.  However, the feature was too broad to be
modeled by the BAL template.  Ly$\gamma$ falls in this  region at
972.5\AA\/.  We see no evidence for a Ly$\gamma$ mini-BAL, so we just
add a broad component; it turns out that the data cannot constrain a
Ly$\gamma$ BAL and mini-BAL simultaneously, as the signal-to-noise
ratio is very low in this region of the spectrum.  The Ly$\gamma$
component allows us to extend the fitting region down to 958\AA\/
before experiencing a poor fit.  We could not find any other transitions
in this region that could significantly add to the opacity.  If the
entire feature down to 955\AA\/originates in Ly$\gamma$ or \ion{C}{3}
alone, and there are no other unidentified transitions, it would imply
that the Ly$\gamma$ or \ion{C}{3} absorption profile is
broader than the \ion{P}{5} profile. This would be surprising if true,
given that these lines are from lower-ionization ions than
\ion{P}{5}.  However, we caution that the evidence for this inference
is rather weak because the signal-to-noise ratio is very low (the
spectrum is almost undetected) and because the region not adequately
fit  by the \ion{C}{3} and Ly$\gamma$ absorption is rather narrow,
only about 4 angstroms wide.  

Finally, we seem to see a \ion{P}{4} mini-BAL and BAL near 950\AA\/, so
we add a component for each of those, fitting down to 940\AA\/.  The
mini-BAL is modeled very well with the template set at the \ion{P}{4}
wavelength. The rest wavelength of Ly$\delta$ is only 0.91\AA\/
below \ion{P}{4}, so it is possible that the absorption feature
originates in hydrogen instead.  Indeed, a comparison of $g_i f_j$
values indicates that the apparent optical depth for Ly$\delta$ should
be at least 18\% that of Ly$\beta$.    However, substituting Ly$\delta$ for
the \ion{P}{4} mini-BAL gives a distinguishably poorer fit
($\chi^2=789$ for the Ly$\delta$ mini-BAL, versus $\chi^2=762.5$ for
the \ion{P}{4} mini-BAL), and a wavelength offset between the model
and the data can be seen.  Thus, while the fit is better for the
mini-BAL in \ion{P}{4}.  This is an odd  result, since the miniBAL is
clearly not present in \ion{P}{5} and \ion{S}{4}, as discussed above.
On the other hand, it is worth reflecting that the signal-to-noise
ratio in this region of the spectrum is very low (Fig.~\ref{fig17}),
and therefore any fit results in this region must be taken with a
grain of salt.   

There are several ions that could contribute to the opacity shortward
of 940\AA\/.  However, the quasar is essentially  undetected in this
region, so no meaningful constraints on the column densities can be
obtained.  

We plot the resulting model and the spectrum in Fig.~\ref{fig9}.  The
results are listed in Table 2.  Except for
the region between 991 and $\sim 1000$\AA\/ in the \ion{O}{6}
absorption feature, and the region between 950 and 960\AA\/ in the
\ion{C}{3}/Ly$\gamma$ absorption feature, the fit is very good.  We
also plot the individual components of the model in Fig.~\ref{fig9}.

\subsubsection{Alternative Deblending}

Fig.~\ref{fig9} shows that we can very nearly successfully fit the
whole {\it FUSE} spectrum using the {\it FUSE} BAL and mini-BAL
templates.  But we cannot be confident about our fit in the \ion{O}{6}
and \ion{C}{3} regions because of the severe blending.  We therefore
consider an alternative extreme case where all of the absorption
between 990 and 1037\AA\/ is due to \ion{O}{6}, and all of the
absorption between 953 and 978\AA\/ is due to \ion{C}{3}.  The spectra
of other BALQSOs often reveal higher velocities in higher ionization
lines \citep[e.g.,][]{jchs01}, so this possibility in the case of
\ion{O}{6} is very plausible.  Thus, we present an alterative
deblending of the blended \ion{O}{6} and \ion{C}{3} regions, as
follows.    We use our successful fits above as a smooth
representation of the data and rederive the optical depths and column
densities for \ion{O}{6} and \ion{C}{3} with this assumption.  Note that
this procedure underestimates the possible \ion{C}{3} optical depth
since the model shown in Fig.~\ref{fig9} does not fully account for the
absorption between $\sim 950$ and $\sim 960$\AA\/.  The net
result is a $\sim 0.4$ dex increase in the column density lower limit
in both ions, compared with the values listed in Table 2.

\subsubsection{Column Density Lower Limits}

Using Eq.\ 9 in \citet{ss91}, we can integrate over the templates to
get estimates of the lower limit of the column densities.  Those
values are given in the fifth column of Table 2.  The uncertainties
given are proportional to the uncertainties in the scale factors in
the fitting, and thus they provide an estimate of only the statistical
uncertainty.    

\section{Discussion}

\subsection{Cloudy Models}

We would like to determine the physical state of the absorbing gas,
and investigate other parameters such as the launching radius
in order to understand outflows in AGN.  While this can be done fairly
robustly in some cases where the signal-to-noise ratio is very good
and the individual lines in multiplets are resolved
\citep[e.g.,][]{arav07}, it is generally difficult in BALs because the
lines are saturated and there is partial covering of the emission
source(s).   But the situation is a little better for 
WPVS~007 because we measure \ion{P}{5} in the {\it FUSE} spectrum, and
it does not seem to be completely saturated.  We find that we can
obtain some interesting and useful limits through analysis of the BALs
using the photoionization and spectral synthesis code {\it Cloudy}
\citep{ferland98}.

\subsubsection{Spectral Energy Distribution}

In this section we try to constrain the properties of the absorbers by
comparing the results of Cloudy models with the column density limits
obtained in \S 3.3.5.  Cloudy requires as input a spectral energy
distribution (SED).  As discussed in \S 2.2, the UV and FUV spectra
are clearly reddened, and the X-rays are generally not detected (until
recently; see below), so it is not possible to build an SED directly
from the WPVS~007 observations.  However, as discussed in \S 2.3, the
emission lines of the NLS1s Mrk~493 and Mrk~335 are very similar to
those of WPVS~007, and these objects have roughly similar optical
luminosity as WPVS~007 ($\log(L_{5500})=39.1$, 39.4, and 40.2 for
WPVS~007, Mrk~493, and Mrk~335, respectively) so as a first
approximation, we assume that the intrinsic SEDs are
similar. Therefore, we construct a SED for WPVS~007 using data from
Mrk~493 and Mrk~335.  

Both Mrk~493 and Mrk~335 have been observed by {\it XMM-Newton} and
therefore simultaneous optical/UV and X-ray data are available.
Mrk~493 was observed once.  During that observation, only the V-band
filter was used.  The Optical Monitor (OM) image clearly shows the
galaxy as well as the bright central AGN, and it is clear that
integration of the flux over the nominal {\it XMM-Newton} extraction
aperture of $12^{\prime\prime}$ will include a significant amount of
galaxy light. Yet we need to use this aperture as that is one that is
calibrated for conversion to energy fluxes.  So we devise an aperture
correction using an observation of the bright quasar PKS~0558$-$504
observed on 2000-05-24.  This object is very luminous, and there is no
apparent galaxy emission in the V-band image.  Using IRAF, we shifted
all 10 images using the V-band filter to a common centroid and summed.
We determined the FWHM of the PSF from the PKS~0558$-$504 image to be
3.34 pixels.  We then extracted the net flux within one FWHM,
considering this to be dominated by the AGN, and within
$12^{\prime\prime}$, using a background annulus between
$14^{\prime\prime}$ and $25^{\prime\prime}$, and derive an aperture
correction of 1.35.  We extracted the flux from the Mrk~493 image in a
region with a radius of one FWHM and also within $12^{\prime\prime}$.
We then correct for deadtime and coincidence losses using the
instructions in the {\it XMM-Newton} Users
Handbook.\footnote{http://heasarc.gsfc.nasa.gov/docs/xmm/uhb/XMM\_UHB.html}.
Correcting the flux from the small aperture using the correction
factor derived from the PKS~0558$-$504 observation, we find that about
70\% of the flux within the large aperture originates in the galaxy.
We also compare the radial profiles from the Mrk~493 and the
PKS~0558$-$504 images, finding that they agree within the size of the
small aperture of 3.34 pixels and verifying extended emission from the
galaxy at larger radii.  Finally, we extract the flux and correct for
the sensitivity degradation using the
instructions.\footnote{http://xmm.vilspa.esa.es/sas/7.0.0/watchout/Evergreen\_tips\_and\_tricks/om\_time\_sensitivity.shtml}.
We then scale the {\it HST} spectrum to match the flux point.  The
{\it XMM-Newton} EPIC data from Mrk~493 and Mrk~335 were analyzed in
the usual way.  Finally each time-averaged spectrum was fitted with a
broken power law between 0.3 and 6 keV. 

There were two {\it XMM-Newton} observations of Mrk~335.  The first
was performed 2000 December 25.  During this observation, images were
obtained using the V, B, U, and UVM2 filters.  The host galaxy
spectrum is expected to be red and therefore to not contribute
significantly to the flux in the UVM2 filter.  Therefore, we use the
results of the standard OM analysis, and then scaled the HST spectrum
to the UVM2 filter flux.  The X-ray data were analyzed in the usual
way except that pileup was present in the data from the PN instrument
and therefore an annular extraction aperture was used.  The spectra
were fitted with a broken power law.  The second observation was 
performed 2006 January 3.  The OM was operated using the UV grism, and
the data were processed in the standard way.  The {\it HST} spectrum
was then scaled to the to grism spectrum.  The X-ray spectrum was
analyzed in the usual way, and fitted to a double broken power law
model.  Finally, the combined UV and X-ray spectra were sampled in
line-free regions to construct SEDs consisting of 10--12 points each.  

The three SEDs are very similar. The $\alpha_{ox}$ values for the
Mrk~493, and first and second Mrk~335 observations were $-1.33$,
$-1.36$, and $-1.32$, respectively.  These are very close to the value
of $-1.22$ predicted for WPVS~007 based on the 5500\AA\/ luminosity
obtained from the {\it HST}  spectrum  and the regression
published by \citet{steffen06} and their cosmological parameters
($H_0=70\rm\, km\,s^{-1}\,Mpc^{-1}$, $\Omega_M=0.3$,
$\Omega_\Lambda=0.7$). Finally, we obtain a merged SED.  Since the {\it
HST} spectrum of Mrk~493 extended to the optical band, we started with
SED for that object.  The Mrk~335 SEDs were rescaled to match the
Mrk~493 SED in the near UV. We found that the Mrk~335 points straddled
the Mrk~493 points in the X-ray, such that an average would have been
very close to the Mrk~493 points.   We found that the FUV of the
Mrk~335 spectra were very slightly higher than the Mrk~493 points, and
appeared to match the optical power law better,  so we replaced the
three far UV points in the SED with the ones from the scaled Mrk~335
spectrum.  We interpolated over the unobservable far UV using a power
law between the {\it HST} and {\it XMM-Newton} PN spectra.  We use the
{\it Cloudy} AGN {\tt kirk} spectrum at higher and lower
frequencies\footnote{The {\it Cloudy}  AGN {\tt  kirk} spectrum is the
{\it Cloudy} AGN spectrum with the parameters $\log T=6.00$
$\alpha_{ox}=-1.40$ $\alpha_{UV}=-0.50$ and
$\alpha_x=-1$. \citep{ferland98}}.  The final spectrum has
$\alpha_{ox}=-1.28$.   

After completing the analysis  outlined above, and associated {\it
  Cloudy} simulations, \citet{grupe08} report the first detection of
hard X-rays in  WPVS~007.  They find evidence for a partially-covered
spectrum, and a plausible deconvolution indicates an intrinsic (i.e.,
unabsorbed) $\alpha_{ox}$ of $-1.9$.  Therefore we construct an
additional continuum spectrum with $\alpha_{ox}=-1.9$ by simply
decreasing the flux of the X-ray and higher energy portion of the SED
developed above  by a factor of 41.3.  The two SEDs are shown in
Fig.~\ref{fig10}.  

\subsubsection{Simulations and Results for the BALs}

Initially we ran {\it  Cloudy}
models for the $\alpha_{ox}=-1.28$ continuum for a range of
parameters: ionization parameter $-3.5\leq \log(U) \leq 0.5$, $\Delta
\log(U)=0.25$; density $8.0 \leq \log(n_H) \leq 12.0$, $\Delta 
\log(n)=0.5$; column density parameter $\log(N_H^{max})$, 
defined as $\log(N_H^{max})=\log(N_H)-\log(U)$, $21.0 \leq
\log(N_H^{max})\leq 24.0$, $\Delta \log(N_H^{max})=0.1$ for a total of
4,743 simulations.  After determining that the absoprtion lines are
not dependent on the gas density \citep[e.g., see also][]{hamann97},
we choose $\log(n)=10.0$ as the nominal density and extend the
simulations to $\log(U)=1.5$.  We also ran the simulations for the
$\alpha_{ox}=-1.9$ continuum for that value of the density.   We use
$\log(N_H^{max})$ rather than just $\log(N_H)$ because a constant
value of $\log(N_H^{max})$ probes similar depths in terms of the
hydrogen ionization front \citep[e.g.,][]{leighly04, clb06, lhjc07}.
We used solar metallicity initially. 

We followed the approach discussed by \citet{arav01} and since used by
a number of other investigators.  Specifically, we compare our
estimates of the column density given in Table 2 with ionic column
densities from the {\it Cloudy} simulations as a function of
ionization parameter and column density.   We emphasize that since the
absorption lines are saturated and partially covered, and since we
estimate the column densities from the apparent optical depths, these
are the {\it lower limits} of the column density.  Most of
the  ionic column densities can be extracted from the ``column
density'' output  from {\it Cloudy}.  There is one relevant 
exception: columns from low-lying excited states of metal ions such as
\ion{S}{4}.  These should be populated at  
typical BLR temperatures and sufficiently high densities according to
their statistical weights $2J+1$.  Thus, the proportion in the ground
($J=3/2$, the 1062\AA\/ component) and fine structure ($J=5/2$; the
1073\AA\/ component) states should be 40\% and 60\% of the column
given by the {\it Cloudy} output.  Interestingly, though, the
estimated column densities given in Table 2 show that the proportions
of the estimated column density upper limits are 44\% and 56\%, very
close to the expected proportions, and therefore comparing the
estimated lower limits with the simulations for both lines would yield
indistinguishable results.  Thus, for the following analysis, we
compare 60\% of the {\it Cloudy} $S^{+3}$ column with the observed
column of the 1073\AA\/ component. 

\begin{figure}
\figurenum{9}
\epsscale{1.0}
\plotone{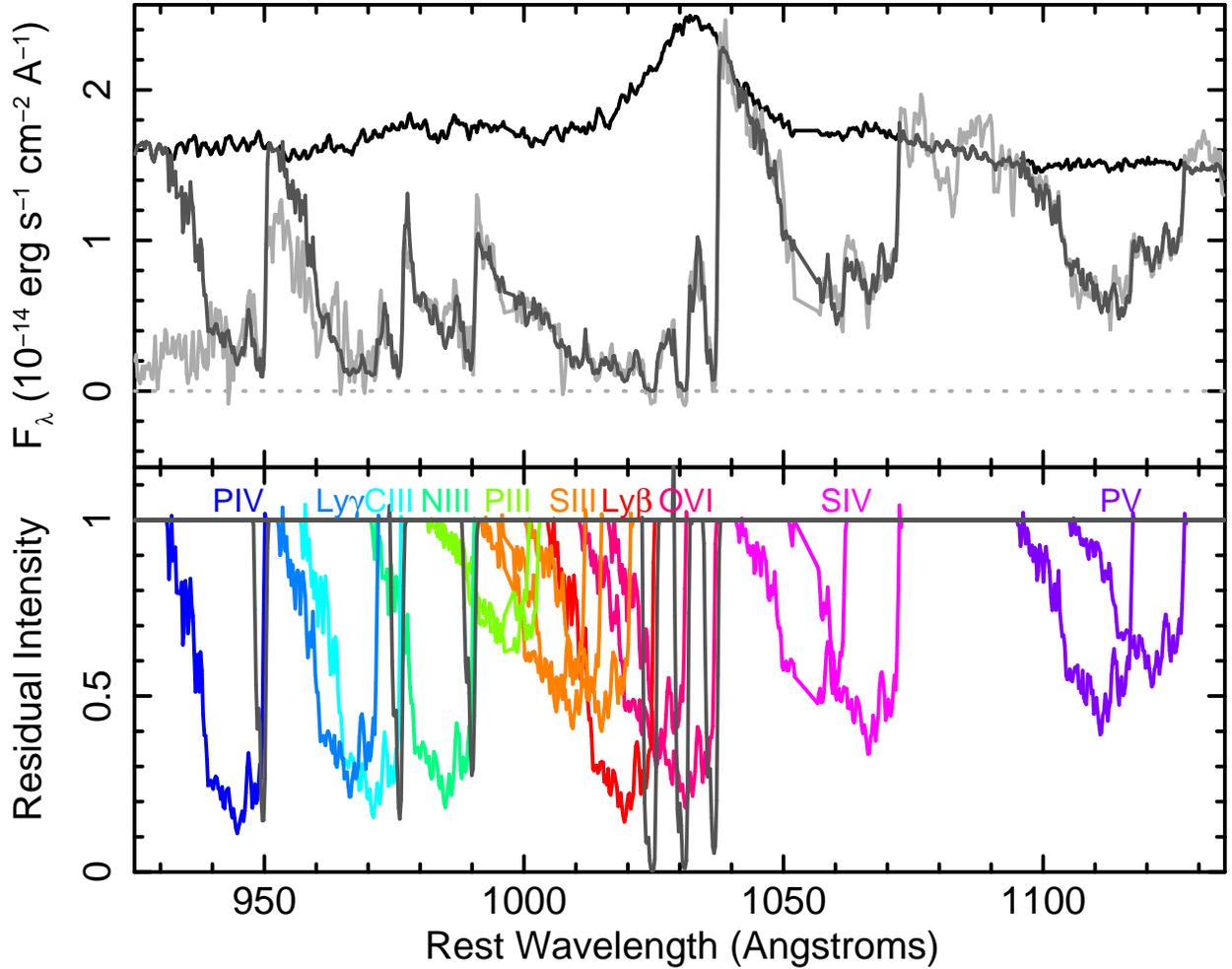}
\caption{The spectral fit derived in \S 3.3.1, 3.3.2, and 3.3.3.  Top
  panel: the 
  light grey line shows the data, the black line shows the assumed 
  continuum, and the dark grey line shows the model fit comprised of
  the BAL template created from \ion{P}{5} (\S 3.1) and the mini-BAL
  template created from \ion{O}{6} and Ly$\beta$ (\S 3.2).  The fit in
  the region of \ion{S}{4} is very good and validates
  the approach.  The fit in the region of \ion{O}{6} is good also, but
  a large number of transitions and ions are needed.  As described in
  \S 3.3.3, we failed to fit the region  between 950 and 958 \AA\/,
  suggesting that the  Ly$\gamma$ or   \ion{C}{3} is broader than the
  \ion{P}{5} BAL, that the assumed continuum level is incorrect and
  should be lower, or   that there are unidentified ions contributing
  to this absorption   feature.   Bottom panel: The   absorption
  features   contributing to the fit shown in the top panel. 
  For the \ion{P}{5} template, the color of the absorption feature
  corresponds to the ion above the panel.  Six incidences of the
  \ion{O}{6} miniBAL are fitted and are shown in grey; from left to
  right, they correspond   to \ion{P}{4}, \ion{C}{3}, \ion{N}{3},
  Ly$\beta$, and two   \ion{O}{6}. \label{fig9}}
\end{figure}

We first examine the results from the $\alpha_{ox}=-1.28$ simulations.
We first consider the BAL deblending that is shown in Fig.~\ref{fig9} using
the columns listed in Table 2.  As mentioned in \S 3.1e also measure
upper limits on the 
column density of a multiplet of \ion{N}{2} near 1085\AA\/, and one of
\ion{Fe}{3} near 1126 (Fig.~\ref{fig9}) by inserting a {\it FUSE} BAL  template
into the model and increasing the normalization until the $\chi^2$
increases by 6.63, corresponding to 99\% confidence for one parameter of
interest. These upper limits are also listed in Table 2.    Fig.~\ref{fig11}
shows the  ionic column contours from Table 2 on the ionization 
parameter/$N_H^{max}$ plane for $\log(n)=10.0$.  Since the column
densities we estimate are lower limits, the solutions should lie
above and to the right of any curve.  Ideally, if the
lines were not saturated, the contours would converge in one area of
the plot, producing   best-fitting values of $\log(U)$ and
$\log(N_H)-\log(U)$.  The presence of \ion{P}{5} implies that lines
of similar ions of more abundant species are saturated
\citep{hamann98}, and therefore we do not expect any region 
of convergence {\it a priori}.  Interestingly, we do find one in which
a majority of the lines converge: for  $\log(U) \geq \sim 0.0$, the
column densities of a number of the intermediate-ionization ions
converge around $\log(N_H)-\log(U)=23.0$.  Notable exceptions are
$O^{+5}$ and $H^0$, for which simulations show  much higher column
density than observed at $\log(U)=0.0$, $\log(N_H)-\log(U)=23.0$, and
S$^{+2}$ and P$^{+2}$, for which simulations show lower column
densities than observed at that point.   If the lines are saturated,
the observed column densities are lower limits, and therefore the
O$^{+5}$ and $H^0$ columns may be consistent with the other lines at
$\log(U)=0.0$, $\log(N_H)-\log(U)=23.0$, but the P$^{+2}$,
and S$^{+2}$ would not be, because the lower limit exceeds the
simulated column densities at that point.  

We note that as expected, there is little dependence on the density
for values in the range $9.5 \leq \log(n) \leq 11$.  The column
densities of P$^{+2}$ are lower at very low densities and increase as
the density is increased, but only for the highest values of
$\log(N_H^{max})$.  At the highest densities, lower-ionization columns
increase, primarily at the highest values of $\log(N_H^{max})$.  In all
cases, the differences are small, and we cannot obtain a markedly
better agreement with the observations by assuming very high or very
low densities. 

As discussed in \S 3.3, while the deblending in terms of the
{\it FUSE} BAL  template presented in Fig.~\ref{fig9} matches the spectrum
well, it is possible that the higher-ionization lines have a broader
profile, and \ion{O}{6} dominates the region between 990\AA\/ and
1037\AA\/, and therefore we presented the alternative deblending in \S
3.3.4.  The alternative deblending does not attribute any measurable
opacity to hydrogen, P$^{+2}$, or S$^{+2}$ (since the onset of
absorption for these lines cannot be clearly seen in the spectrum),
but instead attributes the absorption to O$^{+5}$ between 990\AA\/ and
1037\AA\/, and to C$^{+2}$ between 953 and 978\AA\/. Using these
alternative fitting results, we generate the contour plot as before
(Fig.~\ref{fig11}, right side).  Now we see that all measured ions and
upper limits are consistent near $\log(N_H)-\log(U)=23.0$ and
$\log(U)\ge 0$, except for O$^{+5}$ which is clearly saturated.

We perform the same procedure as above for the simulations using the
$\alpha_{ox}=-1.9$ continuum (Fig.~\ref{fig10} middle panel).  The
results are rather similar in that there is again a region of
convergence for the intermediate-ionization ions, and in this region,
the columns O$^{+5}$ and H$^{0}$ are much larger than our inferred
lower limits, while the columns of P$^{+2}$ and  S$^{+2}$ are much
lower than our inferred lower limits.    The principal difference is
that the location of the convergence of many ionic columns occurs at a
lower value of $N_H^{max}$.  Our nominal best solution for this
continuum is $\log(U)=0.0$ and $\log(N_H)-\log(U)=22.0$.  This result
can be understood by considering that intermediate-ionization ions are
created at a smaller column density in gas illuminated by an
X-ray weak continuum than in gas illuminated by an X-ray normal
continuum because the X-ray weak continuum is unable to produce the
usual higher-ionization ions \citep[e.g.,][]{lhjc07}.  Most 
of the observed absorption lines (except \ion{O}{6}) are from
intermediate-ionization ions.

Quasars are often inferred to have enhanced metallicities
\citep[e.g.,][]{hkfwb02}.  The hydrogen column densities derived
above are large, and as will be shown in \S 4.2, challenge
outflow models.  Inferred hydrogen column densities should be lower if
the metallicity is enhanced.  We therefore try a $\alpha_{ox}=-1.9$
model with  $Z=5$ \citep[that includes helium enhanced by a factor
1.29 and  nitrogen enhanced by a factor of 25;][]{hkfwb02}.   The
results are shown in the lower panel of Fig.~\ref{fig11}.  Here, the
intermediate-ionization ions converge for $\log(U)=0$ at
$\log(N_H)-\log(U)=21.6$.    

Since we expect that partial covering is strongly influencing our
analysis, the fact that the data and models converge for many of the
ions at particular points in these graphs seems somewhat surprising.  
This convergence could be coincidental; alternatively, it could really 
be the case that only \ion{O}{6} and \ion{H}{1} are very saturated and
the intermediate-ionization ions seen in the {\it FUSE} spectrum are
not very saturated.  This idea is supported by the fact that the upper
limits for the undetected absorption lines \ion{Fe}{3} and \ion{N}{2}
are seen in Fig.~\ref{fig11} to lie either in the region of
convergence, or not very far above it.  Therefore, we use the column
densities of convergence at $\log(U)=0$ (specifically,
$\log(N_H)=23.0$, 22.2, and 21.6 for the $\alpha_{ox}=-1.28$ set of
simulations, the $\alpha_{ox}=-1.9$ simulations, and the
$\alpha_{ox}=-1.9$ and $Z=5$ simulations) for discussion in the rest
of the paper, emphasizing that these columns are lower limits.  The
transmitted continua for these cases are shown in 
Fig.~\ref{fig10}.    

Further analysis reveals some interesting facts about these
solutions.  First, the choice of ionization parameter is driven by
$P^{+4}$; specifically,  the ionization parameter must be sufficiently
high to produce a sufficient column of $P^{+4}$ to match the data.
This means that the ions seen in the UV spectra are the tip of
the ion iceberg, since these ions dominate gas with much lower
ionization parameters than  $\log(U)=0$.  That is, for most elements,
most of the atoms are in higher ionization states, including $C^{+3}$
and $N^{+4}$, but also much higher ionization states, indicating that
most of the opacity is in the extreme UV and soft X-ray band.  That
this is true can be see in the absorbed spectra shown in Fig.~\ref{fig10};
much of the continuum in the extreme UV and soft X-ray is absorbed by
gas with this ionization parameter and column density.  

In summary, we find that we can explain the {\it FUSE} spectrum, in
particular, the presence of absorption from the low-abundance element
phosphorus, with a highly ionized ($\log(U) \ge 0.0$), high column
density absorber.  We need to make the assumption, though, that the
high-ionization lines such as \ion{O}{6} are much broader than the
intermediate-ionization line \ion{P}{5}.   Strongly enhanced abundance
of phosphorus is not needed because some lines are saturated, although
they are not black because partial covering is important.  This is
essentially the same result found by \citet{hamann98} for the quasar
PG~1254$+$047 \citep[see also][]{hsjcs01}.   A subtlety explored here
is the dependence of the result on the spectral energy distribution.
The primary distinction for the X-ray weak $\alpha_{ox}=-1.9$
simulations is that the simulated column densities of the
intermediate-ionization ions including P$^{+4}$ are large enough to
match the observed column lower limit at significantly lower values of
$N_H$.   As noted above, this occurs because the X-ray-weak continuum is not able
to create high-ionization ions, and therefore the gas cools via
transitions of  intermediate-ionization ions even in front of the
helium ionization front (the hydrogen ionization front for the 
$\alpha_{ox}=-1.9$ continuum is located beyond $\log(N_H)-\log(U)=24$).
Thus, while partial covering implies that the column densities are
probably larger than inferred from the apparent line optical depths,
and the outflows are still massive enough to challenge radiative-line
driving as the acceleration mechanism  \citep[e.g.,][ see also \S
  4.2]{hsjcs01}, the situation is not as extreme if the continuum is
X-ray weak, because sufficient \ion{P}{5} can be produced at
significantly smaller (factor of 6 for solar abundances, factor of 25
for enhanced abundances) column densities.   

\subsubsection{Implications for the X-ray Spectra}

As discussed in \S 1, WPVS~007 has peculiar X-ray properties.  Its
flux during the {\it ROSAT} All Sky Survey was typical for an AGN of
its luminosity.  Then, many subsequent observations found it to have
nearly disappeared from the X-ray sky.  Could the variable UV absorber be
responsible for the X-ray behavior?   

\begin{figure}
\figurenum{10}
\epsscale{0.4 }
\plotone{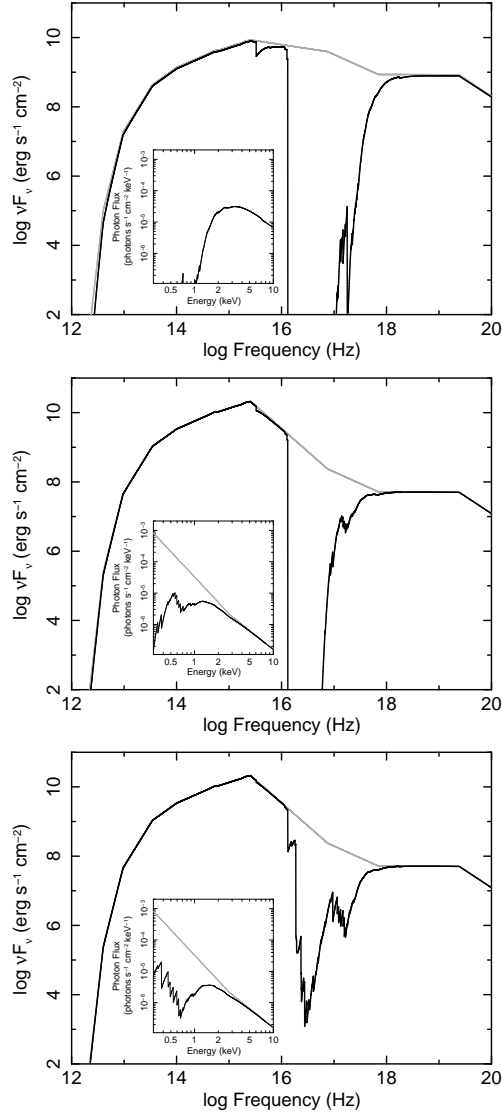}
\caption{Continua used for {\it Cloudy} modeling (see \S 4.1.1; light
  grey) and inferred   absorbed spectra (see \S 4.1.3; black).  The inset
  plots show the X-ray bandpass  photon flux units.  Top: The
  continuum inferred using the {\it HST} and {\it XMM-Newton}
  observations of Mrk~335 and Mrk~493; this continuum has
  $\alpha_{ox}=-1.28$.  The absorbing gas has  $\log(U)=0$,
  $\log(n)=10.0$ and $\log(N_H)=23.0$.  Middle: The X-ray weak
  continuum inferred from the recent {\it Swift} hard X-ray detection
  of WPVS~007 \citep{grupe08}; $\alpha_{ox}=-1.9$.  The absorbing gas
  has $\log(U)=0$, $\log(n)=10.0$, and $\log(N_H)=22.2$.  Bottom: The
  X-ray weak continuum again, but absorbed through metal-rich gas
  ($Z=5$).  The absorbing gas has $\log(U)=0$, $\log(n)=10.0$, and
  $\log(N_H)=21.6$.  \label{fig10}}
\end{figure}

\begin{figure}
\figurenum{11}
\epsscale{0.9}
\plotone{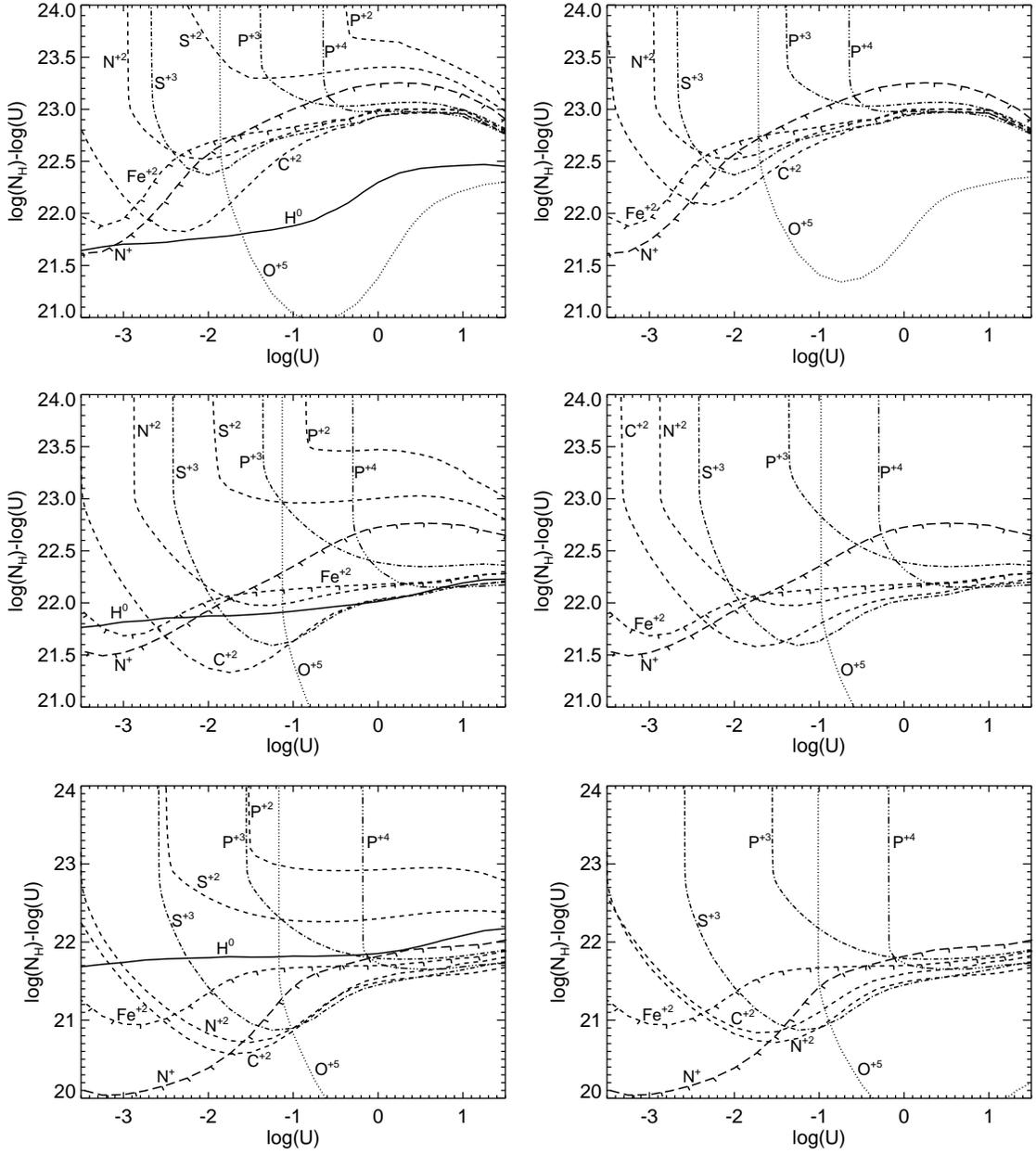}
\caption{Contours of observed column densities on the
  $\log(N_H)-\log(U)$ versus $\log(U)$ plane for $\log(n)=10.0$.
  These column densities were derived using the apparent optical
  depths, implying that they are lower limits and the real solution
  should lie above and to the right of any particular contour.  The
  exceptions are the upper limits for lines that were not observed
  which   are denoted by downward tick marks.    Plots on the left use
  column   densities from the deblending using only   the {\it FUSE}
  BAL   template (and \ion{O}{6}/Ly$\beta$ mini-BAL 
  template) shown in Fig.~\ref{fig9}; plots on the right use column densities
  from the alternative  deblending discussed in \S 3.3.4.  The top,
  middle and bottom panels have the same meaning as in Fig.~\ref{fig10}.  The
  line style shows the ionization state: dotted for +5,
  dot-dot-dot-dash for +4, dot-dash for +3, short dash for +2, long
  dash for +1, and solid for neutral.  \label{fig11}}
\end{figure}

As discussed in \citet{grupe08}, WPVS~007 has been detected
recently twice in X-rays: first, by {\it Chandra}, which observed 10
soft photons from the object \citep[also,][]{vew04}, and more recently
by {\it Swift}, which observed a total of $35.7^{+6.4}_{-6.7}$
photons, including the first hard ($>2\rm\, keV$) X-rays. Moreover, the
{\it Swift} spectrum suggests partial-covering; that is,
the spectrum appears to have a soft component, and a separate absorbed
hard component \citep{grupe08}.  As discussed in \S 
4.1.2 , the column densities and  ionization parameters required to
attain sufficient P$^{+4}$ indicate considerable opacity in the
extreme UV and soft X-rays, as shown in Fig.~\ref{fig10}.  We can use
those absorbed continua to predict the X-ray count rates, as follows.
We first normalize the continua to the blue optical part of the 1995
{\it HST} spectrum, noting that the normalization is somewhat
uncertain since the UV emission has been observed to change by a
factor of 1.5 over the last several years \citep{grupe07, grupe08}.
We convolve the X-ray portion of the  continuum with the Galactic
column density ($N_H(Gal)=2.84\times 10^{20}\rm\, cm^{-2}$) where the
opacity is given analytically by \citet{mm83}.  We fold the results
with the ancillary response files (essentially the effective area or
quantum efficiency curve) generated for the {\it Chandra} and {\it
  Swift} observations. Integrating over the X-ray band pass (taken to
be 0.3--10 keV) yields count rates. Finally, multiplying by the
effective exposure time (9300 seconds for the {\it Chandra}
observation, and 85,508 seconds for the {\it Swift} observation)
yields the number of photons predicted to have been detected.  The
results are given in Table 3. 

These simulations show that, as expected from Fig.~\ref{fig10}, the
predicted absorbed X-ray flux is much lower than the direct flux.  The
largest decrease (factors of 31 and 39 for the {\it Chandra} and {\it
  Swift} observations) is found for the $\alpha_{ox}=-1.28$ continuum,
because the {\it Cloudy} simulations discussed in \S 4.1.2 require the
largest $N_H$ to attain the ion column densities required to produce
the absorption lines.  However, the resulting count rates are much
larger than observed: for example, only 10 photons were observed in
the {\it   Chandra} observation, and these absorbed continua predict
454 photons for the absorbed X-ray spectrum.  Moreover, as seen in
Fig.~\ref{fig10}, these would have been hard X-ray photons ($E > \sim
2\rm\, keV$), while the observed {\it Chandra} photons were all soft
($E < \sim 2 \rm \, keV$).  The same is true for the {\it Swift}
observation: the predicted number of photons is 325, while only 35.7
source photons were observed. 

The correspondence is much better for the $\alpha_{ox}=-1.9$ models.
Again, the predicted absorbed flux is much lower than the direct flux,
but by only by factors of 5--8 for the solar metallicity model, and
7--11 for the $Z=5$ model.  This difference is due to the reduction in
the lower limit of the column density required to produce the FUV line
opacity, as discussed in \S 4.1.2.  Interestingly, the $Z=5$
metallicity model requires a smaller column density, yet predicts a
lower X-ray flux.  This is due to the increased opacity in the soft
X-rays from the additional metals.  The count rates are still somewhat
too high; for example, the predicted {\it Swift} spectrum has 70
counts for the $Z=5$ model, and only 35.7 were observed.  But
considering the uncertainty on the absolute normalization and the fact
that the derived columns are upper limits, the agreement within a
factor of two is remarkably good.  

Finally, it should be noted that the predicted X-ray spectra  shown in
the lower two panels of Fig.~\ref{fig10} are hard, absorbed spectra
but with a recovery toward soft X-rays that would be able to be fitted
by partial covering models.  Thus, these models would correspond well
to the  {\it Swift} spectrum \citep{grupe08} but would not explain the
{\it Chandra} spectrum in which no hard X-rays were observed.

These estimates show that if the intrinsic $\alpha_{ox}$ is $-1.28$,
the gas responsible for the UV absorption lines would have been
insufficient to explain the X-ray  weakness by absorption in the same
gas.  However, if the intrinsic $\alpha_{ox}$ is $-1.9$, as indicated
by the {\it Swift} spectrum, the same gas could have been responsible
for the UV absorption lines, the absorbed component of the X-ray
spectrum, and also potentially the X-ray soft excess.  There are many
caveats, however.  The primary one is that the {\it FUSE} and {\it
  Swift} observations were separated by four years, and since this is
clearly an evolving system, there is no guarantee that the X-ray and
far UV spectra were the same as observed.  In addition, the \ion{P}{5}
absorption profile shows signs of saturation plus partial covering,
implying that the real column densities are larger than the lower
limits discussed here. Furthermore, while the idea that the UV and
X-ray absorption occur in the same gas is simple and attractive, there
is no physical requirement that that be the case, since the X-ray and
UV emission may not be produced in the same region of the central
engine \citep[e.g.,][]{se07}. 

\subsubsection{Simulations and Results for the MiniBALs}

\begin{figure}
\figurenum{12}
\epsscale{0.5}
\plotone{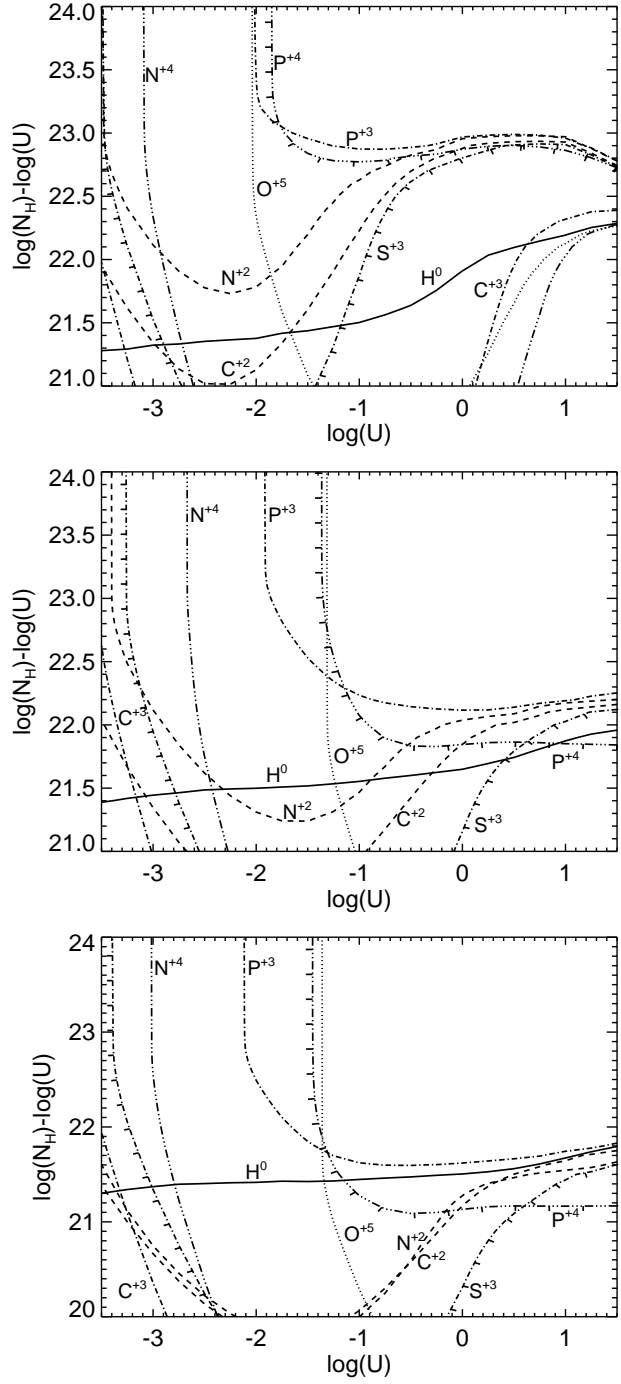}
\caption{Contours of observed column densities on the
  $\log(N_H)-\log(U)$ versus $\log(U)$ plane for $\log(n)=10.0$ for
  the mini-BALs.  Lines have the same meaning as in Fig.~\ref{fig11}.  \label{fig12}}
\end{figure}

We next consider the mini-BALs.  We combine the results for the {\it HST}
and {\it FUSE} observations, noting that these were not simultaneous
and it is possible that the apparent optical depths changed between
the two observations.  Following the same procedure as above, we
produce the contour plot shown in Fig.~\ref{fig12}, showing again the
results for three cases: $\alpha_{ox}=-1.28$, $\alpha_{ox}=-1.9$, and
$\alpha_{ox}=-1.9$ with  $Z=5$.  An important constraint on the
mini-BALs is that we  do not detect the mini-BALs in either P$^{+4}$
or S$^{+3}$ (Fig.~\ref{fig8}).  We obtain upper limits on the mini-BAL
columns of these ions by including the mini-BAL template in the
spectral-fitting model, and increasing its opacity until the $\chi^2$
increases by 6.63, corresponding to 99\% confidence for one parameter
of interest. Those values are listed in Table 2.

Considering first the $\alpha_{ox}=-1.28$ continuum, we find that we
obtain a reasonable solution for $\log(U) \geq \approx -0.3$ and a
$\log(N_H)-\log(U)$ lower limit of 22.8, corresponding to a
$\log(N_H)$ lower limit of 22.5.  This solution is consistent with the
upper limit on mini-BAL column densities of both P$^{+4}$  and
S$^{+3}$.  This solution indicates that the Ly$\alpha$, \ion{C}{4} and
\ion{N}{5} absorption lines are strongly saturated.

We next present the results for $\alpha_{ox}=-1.9$ (middle panel of
Fig.~\ref{fig12}).  We find a rough correspondence for $\log(U) \geq
0.5$ and $\log(N_H)-\log(U)$ lower limit of 22.2 corresponding to a
$\log(N_H)$ lower limit of 22.8.   Again, Ly$\alpha$, \ion{C}{4} and
\ion{N}{5} absorption lines are strongly saturated, consistent with
the observed absorption line intensity ratios (\S  2.3).  However,
this solution is not consistent with the upper limit on mini-BAL
column densities of both P$^{+4}$  and S$^{+3}$. 

Finally, for $\alpha_{ox}=-1.9$ and $Z=5$, we find   a rough
correspondence for $\log(U) \geq 0.5$ and $\log(N_H)-\log(U)$ lower
limit of 21.5 corresponding to a $\log(N_H)$ lower limit of 22.0.  In
this solution, however, the column densities of P$^{+4}$  and
S$^{+3}$ would exceed the upper limits.  Thus, it does not appear to
be possible to produce the miniBALs and upper limits with either of
the two continua we are considering.  

It may be possible to reconcile the column density constraints from
the detections and the upper limits if the upper limits on the
unobserved lines are too low.  As discussed in \citet{hamann01}, this
could be true if the covering factor is higher for strong/detected
lines than for weak lines.  In a patchy absorber with a range of
optical depths \citep[see, e.g., Fig.\ 6 in][]{hamann01}, strong lines
could have large optical depth over a larger area, while weak ones
could still have $\tau >1$ but over a smaller area, leading to a lower
observed equivalent width.  Our assumption of complete coverage for
all lines would then lead to a too-low upper limit on the undetected
weak lines. 

\subsection{Luminosity Dependence and Launch Radius}

As discussed in \S 1, outflows are common in AGN, but their nature 
differs between high and low luminosity objects.  High velocity
outflows are generally limited to high luminosity objects, while lower
luminosity objects have typical outflow velocities of only $<10^{3}\rm
\, km\, s^{-1}$.  The dependence of absorption on other AGN parameters
has been systematically investigated using a sample of low-redshift
($z<0.5$) quasars with $M_V$ between $\sim -21$ and $\sim -27$ by
\citet{blw00} and \citet{lb02}.  A primary result of \citet{blw00} is
that there exists a significant correlation between $\alpha_{ox}$ and
\ion{C}{4} absorption-line  equivalent width, suggesting that the
primary cause of X-ray weakness is absorption, with a continuum of
absorption columns connecting unabsorbed objects to BALQSOs.  In
\citet{lb02} ideas associated with that continuum of absorption
columns  were further developed; in particular, the question of what
makes a soft X-ray weak quasar a BALQSO was addressed.  WPVS~007 is
soft X-ray weak, both as a consequence of absorption, and it is
apparently also intrinsically soft X-ray weak \citep{grupe08}.  As
shown in this paper, it had broad absorption lines during the {\it
  FUSE} observation in 2003.  In this section, we compare some of our
results with those of \citet{blw00} and \citet{lb02}. 

One of the principal results of \citet{lb02} is that there is a strong
dependence of outflow properties on luminosity.  Specifically, there
are strong positive correlations between the \ion{C}{4} equivalent
width and $V_{max}$, the outflow maximum velocity, with the optical
luminosity $M_V$ for soft X-ray weak quasars, and at any luminosity,
soft X-ray weak quasars had the largest equivalent widths and maximum
velocities.  As discussed by \citet{lb02}, this behavior is consistent
with outflow scenarios for outflows driven by either dust or line
opacity.  

\begin{figure}
\figurenum{13}
\epsscale{1.0}
\plotone{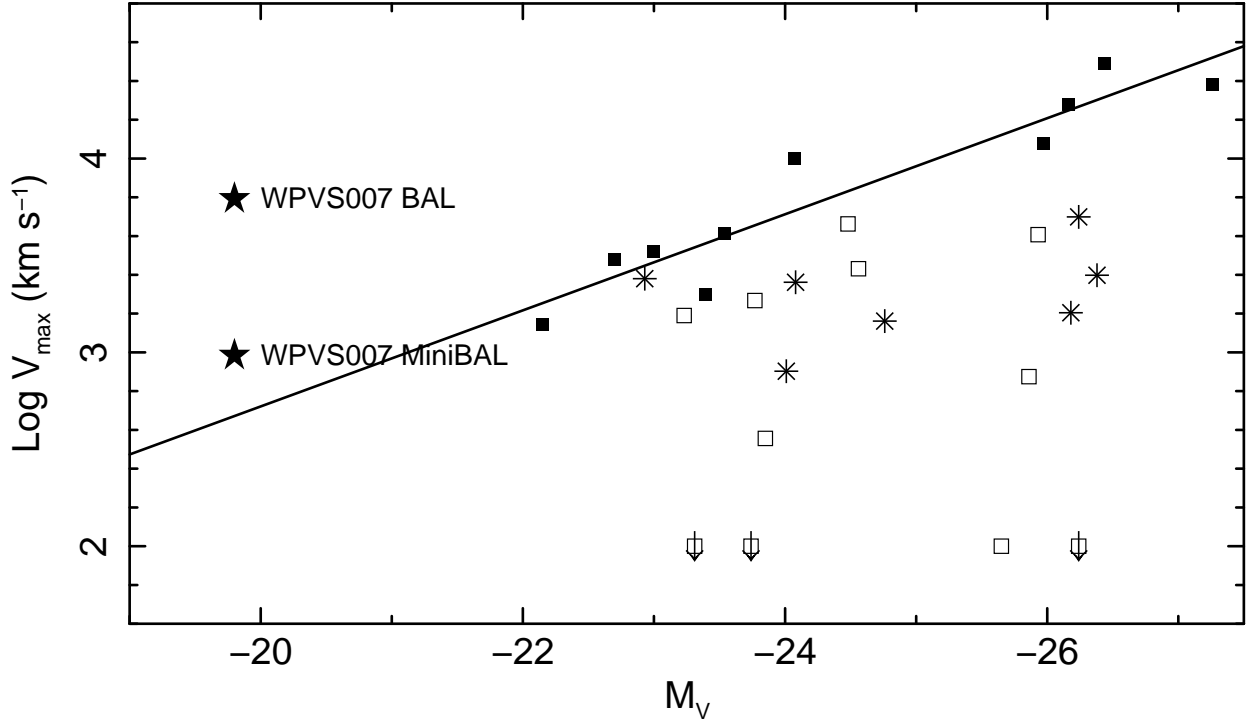}
\caption{WPVS~007 in comparison with low-redshift quasars presented by
  \citet{lb02}.  This is an adoption of their Fig.\ 6 and the data was
  taken from their Table 1. The symbols are taken from their paper
  with the insubstantial modification that we do not differentiate
  between objects with data in \ion{C}{4} and objects with data in
  Ly$\alpha$, as follows: filled squares are soft X-ray weak quasars
  (SXWQs),   asteristics are non-SXWQs with intermediate absorption
  (1\AA\/ $<$ EW $<$ 10\AA\/), open squares are AGN with weak
  absorption (EW$<$ 1\AA\/), and arrows are objects with
  $V_{max}<10\rm \, km\, s^{-1}$.  The solid line is the regression to
  the SXWQ data presented by \citet{lb02}. In addition, we plot our
  results from the {\it HST} and {\it FUSE} observation of WPVS~007.
  The filled star marked ``WPVS~007 BAL'' shows the $V_{max}$ from the
  {\it FUSE} BAL template derivation presented in \S 3.1.  It lies far
  above the regression, indicating a rather large maximum velocity for
  its luminosity.  The filled star marked ``WPVS~007 MiniBAL'' shows
  the mean of the $V_{max}$ for the {\it HST} and {\it FUSE} mini-BALs
  derived in \S 2.3 and 3.2, respectively.   The mini-BAL $V_{max}$ is
  more consistent with that expected of an object of WPVS~007's
  luminosity.   \label{fig13}}
\end{figure}

We compare WPVS~007 with the quasar sample from \citet{lb02} in
Fig.~\ref{fig13}.  The $M_V$ value of $-19.8$ was derived from the
dereddened, restframe {\it HST} spectrum using $H_0=50\rm\, km\,s^{-1}\,
Mpc^{-1}$ and $q_0=0.1$ so as to be consistent with the other data.   
 The straight line in that graph shows the best fit to the soft
X-ray quiet quasars given by \citet{lb02}; they note that it suggests
an upper envelope as would be expected from radiation-driven winds.
The filled star labeled WPVS~007 BAL is taken from the {\it FUSE} BAL 
template derived in \S 3.3. The $V_{max}=6179\rm\,km\,s^{-1}$ for that
absorption line is seen to be a factor of 13 in excess of the value
expected if the regression holds for low luminosity objects.  We note
that most of the data shown in Fig.~\ref{fig13} were derived from \ion{C}{4}
absorption lines, and it is not clear that the \ion{P}{5} absorption
line should have the same profile.  However, as shown by
\citet{jchs01}, \ion{P}{5} absorption is typically better fit by
templates derived from \ion{Si}{4}, and templates from \ion{Si}{4}
tend to be narrower than templates from \ion{C}{4}.  This evidence,
plus the fact that we infer that the  \ion{O}{6} absorption profile in
WPVS~007 is most likely significantly broader than the \ion{P}{5}
profile (\S 3.3.2 \& 4.1.2), indicates that the point shown in
Fig.~\ref{fig13} for the 
BAL may be a lower limit, and the discrepancy between that point and
the \citet{lb02} regression may be even larger.  We also plot the
$V_{max}$ for the mini-BALs, where we show the mean of the values  for
the {\it HST} mini-BAL (\S 2.3) and the {\it FUSE} mini-BAL (\S 3.2).
That $V_{max}$ is very near the \citet{lb02} regression, indicating
the outflow maximum velocity that might be expected for this
relatively low-luminosity object. 

Such a large maximum velocity in a low-luminosity object such as
WPVS~007 may cause problems for acceleration models.  In a
radiative-line-driving scenario, fundamental limits can be placed by
simply considering $F=ma$ where the force of the
radiative line driving, essentially turning continuum luminosity into
wind momentum, is opposed by the force of gravity due to the black
hole.  The $F=ma$ equation, with the acceleration $a=v(dv/dr)$
integrated to get the terminal velocity,  and converted to parameters 
suitable for this situation, is presented as Equation 3 in
\citet{hamann98} \citep[also,][]{hsjcs01}: 

$$V_{\infty} \approx 9300 R_{0.1}^{-1/2} \large( \frac{f_{0.1}
  L_{46}}{N_{22}} - 0.1M_8 \large)^{1/2}$$ 

where $R_{0.1}$ is the launch radius in units of $0.1\rm\, pc$,
$f_{0.1}$ is the fraction of the total flux incident on the flow that
is absorbed or scattered by the wind,
relative to 10\%, $L_{46}$ is the luminosity in terms of $10^{46}\rm\,
erg\, s^{-1}$, $N_{22}$ is the column density relative to $10^{22}\rm\,
cm^{-2}$, and $M_8$ is the black hole mass in terms of $10^8 \rm
M_\odot$.  

To use the \citet{hamann98} equation, we need an estimate of the black
hole mass.  We estimate the black hole mass using standard methods
using the {\it HST} spectrum.  We model the region including H$\beta$
with a linear continuum, an \ion{Fe}{2} template, a Lorentzian profile
for broad H$\beta$, and two Gaussians for \ion{O}{3} constrained to
have flux ratios of three to one, equal width, and fixed separation.
We also use a Gaussian for the NLR component of H$\beta$, fixing the
flux to be one tenth that of [\ion{O}{3}]$\, \lambda 5007$
\citep{cohen83, leighly99}.  The [\ion{O}{3}] lines are slightly
blueshifted with respect to the peak of H$\beta$, as is sometimes
found in Narrow-line Seyfert 1 galaxies \citep[e.g.,][]{ako05,byz05},
and relative to lower-ionization narrow-line region lines, and
therefore we fix wavelength of the narrow component of H$\beta$ to
match that of broad H$\beta$. The resulting fit is very good; e.g.,
substituting a Gaussian for the Lorentzian to model broad H$\beta$
yields a much worse fit.  The width of H$\beta$ is measured to be
$1190\rm\, km\,s^{-1}$.  We obtain the rest frame flux at 5100\AA\/
from the {\it HST} spectrum ($1.1\times 10^{-14} \rm\, erg\, s^{-1}\,
cm^{-2}\, $\AA\/$^{-1}$).  We compute the broad-line region radius
using the regressions found by \citet{bentz06} using the flux and a
luminosity distance of $126.2 \rm \, Mpc$ using their (inferred)
cosmological parameters of $H_0=70\rm\, km\, s^{-1}\, Mpc^{-1}$,
$\Omega_M=0.27$, and $\Omega_\Lambda=0.73$. That is found to be
$\log(R_{BLR})=1.062$ in units of light-days. Finally, we compute the
disperson of the H$\beta$ line profile obtained after subtracting the
other fitted components, and referring to \citet{collin06}, we use a
scale factor of 1.2 to yield a black hole mass of $4.1 \times 10^6 \rm
\, M_\odot$.

We estimate the bolometric luminosity by integrating over the
{\it Cloudy} continuum spectra discussed in \S 4.1.1 after they had
been   normalized to match the rest-frame optical spectrum (recall, as
discussed in \S 2.2, the UV continuum is heavily reddened).  These
are given in Table 4.  We obtain a rough estimate of the fraction of
the continuum absorbed in the wind by integrating over the absorbed
continuum and comparing with the unabsorbed value.  This includes only
thermal velocity for the lines, so it provides a lower limit on the
fraction of the continuum able to accelerate the wind.   These
estimates vary from $f_{0.1}=4.1$ for the $\alpha_{ox}=-1.28$
continuum, from which we infer the highest column density, to
$f_{0.1}=0.67$ for the $\alpha_{ox}=-1.9$, $Z=1$ case and
$f_{0.1}=0.44$ for the $\alpha_{ox}=-1.9$, $Z=5$ case, from which we
require lower column densities, as discussed in \S 4.1.2. We
emphasize again, though, that only lower limits on the column
densities were obtained.   Finally, using the opacity profile  
in Fig.~\ref{fig4}, we conservatively estimate a terminal velocity of
about $4,000\rm\, km\,s^{-1}$; that is, we observe measurable opacity to
about $6,000\rm\, km\,s^{-1}$, but the opacity has decreased
significantly from the maximum by $4,000\rm\, km\,s^{-1}$.  It should
be noted, however, that as discussed in previously in this section,
and in \S 3.3.4, the \ion{O}{6} absorption line may have a much higher
terminal velocity, up to $12,000\rm \, km\, s^{-1}$.  

The estimate of the launch radii for the $\alpha_{ox}=-1.28$ model,
the $\alpha_{ox}=-1.9$ model, and the $Z=5$ $\alpha_{ox}=-1.9$ model
are given in Table 4.  For the $\alpha_{ox}=-1.9$, $Z=1$ model, we
obtain a negative radius, which means that the momentum of the
absorbed continuum is not sufficient to counteract gravity, and an
outflow would not be predicted.  The largest outflow radius is found
for the $\alpha_{ox}=-1.9$, $Z=5$ model, of $R_{0.1}=0.0070$, or
$7\times 10^{-4}\rm\, pc$, or $2.16\times 10^{15}\rm\, cm$. A
$4.1\times 10^6\rm\, M_\odot$ black hole has a Schwarzschild radius of
$R_S=1.2\times 10^{12}$, implying that the launch radius is on the
order of $1780 \rm R_S$.   

The radius of the broad line region was estimated above to be
$\log(R_{BLR})=1.062$, corresponding to $3\times 10^{16}\rm\, cm$,
about 14 times the inferred launch radius.  This is could be a problem
for the radiative-line-driving model, since the broad absorption-line
region is thought to lie at a larger radius than the broad emission-line
region.  This may imply that a magnetic component of acceleration is
necessary \citep[e.g.,][]{bottorff97,everett05}.  On the other hand,
the radius of the observed BAL gas may be much larger than the launch
radius, and the BAL simply does not intercept the line of sight until
it reaches the BLR or beyond.  

\subsection{Absorption Variability in WPVS~007}

Variability in broad absorption lines in quasars is common.  Usually,
changes in apparent optical depth, coupling real changes in opacity
and covering fraction, rather than changes in velocity profile are
observed.  Recently, more dramatic changes have been reported
\citep[e.g.,][and references therein]{lundgren07, gibson08}, although
it is important to note that these variability studies were limited to
known BALs.  The BAL in WPVS~007 is distinct in several ways.  First,
it is one of the very few known cases in which a BAL flow appeared;
other possibly similar examples have been found in the quasars
TEX~1726$+$344 \citep{ma02} and J105400.40$+$034801.2
\citep{hamann08}.  Second, as shown in Fig.~\ref{fig13}, it has quite
a low luminosity for the maximum velocity of the outflow.    

The development of the BAL in WPVS~007 may be associated with its low
luminosity.  WPVS~007 has a small black hole mass and correspondingly
small central engine, emission- and absorption-line regions.  For
example, LBQS~1212$+$1445, the comparison object shown in Fig.~\ref{fig6},
has an outflow with similar maximum velocity, but with $M_V=-27.6$, it
is 100 times more luminous, and therefore the emission regions are
expected to be 10 times larger.  

Currently, we have only two epochs of UV observations of WPVS~007: one
with only mini-BALs (the 1996 {\it HST} observation), and one with both
mini-BALs and BALs (the 2003 {\it FUSE} observation.  Thus, we can only
speculate about the origin of the variability.  In other variable
objects, a favored explanation is generally a change in covering
fraction.  Thus, perhaps WPVS~007 always had a BAL outflow, but
rotation of the accretion disk (assuming the BAL arises from a disk
wind) may have simply rotated it into the field of view.  This 
argument is used to explain variability in quasar-luminosity BALQSOs
in \citet{gibson08} over 3--6 years, and thus it would certainly be a
viable explanation for change in WPVS~007 with its lower luminosity,
thus, smaller size- and time-scales, and longer interval between
observations.  

Another, perhaps more exciting possibility is that the BAL outflow
developed over the time scale of seven years.  The observations were
separated by $2.22\times 10^8\rm \, s$ in the rest frame.  If the
velocity of the bulk of the outflow was $4,000\rm \, km\, s^{-1}$,
then it could have covered a distance of $8.9 \times 10^{16}\rm \, cm$
in the interval between the observations.  As discussed by
\citet{hamann08} and references therein, the high-ionization
absorption plausibly occurs just outside the corresponding broad
emission line radius.   We estimate the size of the
BLR in WPVS~007 to be $3\times 10^{16}\rm \, cm$, and thus there would
have been just sufficient time for the absorbing flow to cover the
BLR.  In an object 100 times more luminous, such as LBQS~1212$+$1445,
and similar outflow velocity, it would have taken ten times longer, or
about 70 years.  So WPVS~007 may be unique in that we observed the
development of a BAL.  Since we can observe such extreme evolution on
human time scales, WPVS~007 is an important object for understanding
BAL winds physical conditions and driving mechanisms.  

This argument is compelling, but cannot simply explain both the X-ray
and UV variability.  WPVS~007 was already X-ray weak by the time of
the {\it HST} observation in which only the miniBAL was observed.  So
the emergence of the UV BAL into the line of sight by the {\it FUSE}
observation could not have been directly associated with the X-ray
absorption.  Rather, it is possible that the X-ray absorption
\citep[perhaps a separate but associated component of ``shielding
gas''][]{se07} moved into the line of sight some time after the {\it 
ROSAT} All Sky Survey observation, and before the {\it HST} and {\it
FUSE} observations.  On the other hand, as shown in \S 4.1.3, the gas
that produces the UV BALs can nicely explain the {\it Swift} X-ray
spectrum.  Since we have only two epochs of UV spectroscopicdata, with
no simultaneous X-ray coverage, it is difficult to determine precisely
what happened.     

Clearly, we need more observations to understand what is happening in
WPVS~007.  These are now even more urgent with the discovery from {\it
Swift} that the absorption in the X-ray band is apparently changing.
A second {\it FUSE} observation was approved, and even though
WPVS~007, with a declination of $-51$ was in the region of the sky
that could be observed after the loss of the reaction wheels, the
observation was never scheduled before the satellite was
decommissioned. An observation using {\it HST} COS has been approved, 
along with a contemporaneous {\it Chandra} observation, and therefore
we still have a chance of observing further absorption evolution of this
interesting object.  

\section{Summary}

We present optical and UV observations of the unusual transient AGN
WPVS~007.  This Narrow-line Seyfert 1 galaxy was observed to be as
bright as an average AGN of its luminosity in the {\it ROSAT} All Sky
Survey, but then nearly disappeared from the X-ray sky in subsequent
observation.  We present a reanalysis of the 1996 {\it HST} optical
and UV spectrum, and an analysis of the 2003 {\it FUSE} observation.
The principle results follow. 

\begin{itemize}
\item  We discovered the emergence of broad absorption line  features
  between the {\it   HST} and {\it FUSE} observations.  In 
  the {\it HST} observation, mini-BALs with $v_{max} \sim 900\rm \,
  km\, s^{-1}$ and $FWHM \sim 550 \rm \, km\, s^{-1}$ were observed.
  In the {\it FUSE} observation, the mini-BALs were still present, and
  an additional BAL component with  $v_{max} \sim 6000 \rm\,  km\,
  s^{-1}$ and $FWHM \sim 3400 \rm \, km\, s^{-1}$ had appeared.  While  
  variability of absorption lines in BAL quasars and Seyfert galaxies
  is frequently seen, it is usually limited to changes in the apparent
  optical depth of the line.  The change in the line strength
  described here   is the  most dramatic ever observed in an AGN.

\item Using a template method of analysis, we obtain the apparent
  optical depths of the absorption lines, and derive corresponding
  ionic column densities for both the BALs seen in the {\it FUSE}
  spectrum,  and the mini-BALs seen in both the {\it HST} and {\it
  FUSE} spectra.   BALs are thought to be saturated and have
  potentially velocity-dependent partial covering, so the measured
  ionic column densities are lower limits.    We then use {\it Cloudy}
  to try to obtain some information about the physical conditions of
  the absorbing gas.  We use two different continua: one with
  $\alpha_{ox}=-1.28$, similar to that of a typical quasar with the same
  optical luminosity  as WPVS~007, and one with $\alpha_{ox}=-1.9$,
  corresponding to the inferred intrinsic value from the recent hard
  X-ray detection by {\it Swift}   \citep{grupe08}.  For the BALs, we
  find that \ion{P}{5} constrains   the column density and ionization
  parameter.  For the   $\alpha_{ox}=-1.28$ continuum, we find that
  $\log(U)\geq 0$, and $\log(N_H) \geq 23$.  For the
  $\alpha_{ox}=-1.9$ continuum, we obtain approximately the same limit
  on the   ionization  parameter, but require $\log(N_H) \geq 22.2$.
  For the   $\alpha_{ox}=-1.9$ continuum and $Z=5$ metallicity, the
  column   density lower limit  becomes $\log(N_H) \geq 21.6$.  The
  total column density is lower for the steeper  (intrinsically X-ray
  weak)  continuum because  intermediate ionization ions (in
  particular P$^{+4}$) are produced at smaller column densities in gas 
  illuminated by a steep SED compared with gas illuminated by a flat
  SED   \citep[e.g.,][]{lhjc07}.  These  large column density
  estimates are similar to those obtained  previously for   the
  \ion{P}{5} quasar PG~1254$+$047   \citep{hamann98}.    Acceleration
  of these large   column densities   challenge radiative-line driving
  as a mechanism.     We also point out that since the observed
  \ion{P}{5} absorption can   be explained by lower total column
  densities using an intrinsically   X-ray weak spectrum, we might
  expect to find \ion{P}{5} absorption   preferentially more often (or
  stronger) in quasars that are   intrinsically X-ray weak. 

\item  The high ionization parameters and high column densities
  inferred for the BALs predict X-ray absorption, implying that 
  BALQSOs should be observed to be soft X-ray weak, as they are 
  \citep[e.g.,][]{blw00}.  In addition, a recent long {\it Swift}
  observation indicates that WPVS~007 is also intrinsically X-ray
  weak, and thus there are apparently two effects leading to the
  long-term X-ray faintness of WPVS~007.  We find that for the X-ray
  normal $\alpha_{ox}=-1.28$ SED, the X-ray absorption is insufficient
  to  explain the low {\it Chandra} and {\it Swift} count rates.
  However,  for the intrinsically X-ray weak $\alpha_{ox}=-1.9$ SED,
  with or without enhanced abundances, the inferred absorption yields 
  predicted count rates are only slightly higher than observed. 
  The predicted X-ray spectra for $\alpha_{ox}=-1.9$ have a heavily
  absorbed   hard component, and a   soft component due to reduced
  opacity resulting from the high ionization parameter.  Thus, they
  resemble the spectrum from the recent {\it Swift} observation, but
  cannot explain the {\it   Chandra} observation in which only soft
  photons were detected \citep{grupe08}.  It is possible that the
  absorption column may have been larger  during the  {\it   Chandra}
  observation.  Despite the fact that the   {\it FUSE} and {\it
    Swift} observations were separated by four years, this consistency
  supports the idea that the X-ray weakness and the broad absorption
  lines both result from absorption in the same gas.   The weakness in
  this argument is that WPVS~007 was already X-ray weak during the
  {\it HST} observation, before the UV BAL emerged.  

\item Given the luminosity of the object, an estimate of its black
  hole mass, the BAL terminal velocity
  $V_{max}$, the lower limit on the absorption column, and the
  fraction of the  bolometric luminosity inferred to be absorbed in
  the outflow, we can estimate the launch radius for the outflow
  using Eq.\ 3 from   \citet{hamann98}.  We find a negative radius for
  the   $\alpha_{ox}=-1.9$ solar abundance result, indicating that
  there  is   insufficient momentum in the absorbed photons to
  accelerate the gas   to the observed terminal velocity.  The 
  largest launch radius ($2.2\times 10^{15} \rm\,  cm$) was obtained
  from the  $\alpha_{ox}=-1.9$, $Z=5$ model, which predicted the
  lowest estimated column density.  However, this launch radius is a
  fraction (1/14) of the estimated size of the broad-line region,
  implying that we must not see it at the launch radius, but rather it
  crosses our line of sight further downstream.  On the other hand, if 
  the BAL outflows originate at larger   radius than the BELR, it
  appears that  radiative line   driving is insufficient to accelerate
  the outflow,   and a form of magnetic driving may be   necessary. 

\item  All other known cases of broad absorption lines are in luminous
  quasars.  WPVS~007 has a uniquely low luminosity compared with other
  objects with similar $V_{max}$.  It is therefore a significant
  outlier on the $M_V/V_{max}$ relationship  \citep[Fig.\ 13,][]{lb02}.  

\item Given that there are only two epochs of UV spectroscopic
  observations of   WPVS~007, one without and one with the BAL, it is
  impossible to   determine the nature of the absorption variability.
  It may be a   change in covering fraction due to a wind from an
  accretion disk   orbiting into the line of sight, as has been
  suggested to explain   variability in other BALs \citep{gibson08}.
  But given WPVS~007's   low luminosity, quite unusual for an object
  with BAL $v_{max} \sim   6000\, \rm km\, s^{-1}$, and
  correspondingly small size scales, it   is possible that we have
  observed the development and onset of the   BAL outflow.  An
  approved {\it HST} observation using COS may help   us understand
  the variability evolution in this unusual object. 

\end{itemize}



\acknowledgments

KML is grateful for useful discussion with Mike Crenshaw, Xinyu Dai,
and Steve Kraemer.  We are also grateful that Mike Crenshaw, knowing
that we were working on this paper, opted not to include it in the
compilation of {\it FUSE} spectra from AGN and quasars \citep{dunn07}.
Part of the work presented here was done while KML was on sabbatical
at the Department of Astronomy at The Ohio State University, and she
thanks the members of the department for their hospitality.  OU
graduate student Hemantha Maddumage and 2006 OU REU student Curtis
McCully performed the reduction of the {XMM-Newton} EPIC data from
Mrk~493 and Mrk~335.  This research was supported by NASA contracts
NAG5-13701 (K.M.L.\ \& D.C.)  and NNX07AH67G (D.G.).  The authors also
thank an anonymous referee who's comments lead to greater clarity in
the paper.  Some of the data presented in this paper were obtained
from the Multimission Archive at the Space Telescope Science Institute
(MAST).  STScI is operated by the Association of Universities for
Research in Astronomy, Inc., under NASA contract NAS5-26555.  Support
for MAST for non-HST data is provided by the NASA Office of Space
Science via grant NAG5-7584 and by other grants and contracts.  This
research has made use of the NASA/IPAC Extragalactic Database (NED)
which is operated by the Jet Propulsion Laboratory, California
Institute of Technology, under contract with the National Aeronautics
and Space Administration. This research has made use of data obtained
from the High Energy Astrophysics Science Archive Research Center
(HEASARC), provided by NASA's Goddard Space Flight Center.  KML and
DAC gratefully acknowledge support by NASA grant NAG5-13701 (FUSE).



{\it Facilities:} \facility{FUSE}, \facility{HST (FOS)}



\appendix

\section{FUSE Data Processing and Background Subtraction}

The data were processed with CalFUSE version 3.1.8.  The standard
procedure circa summer 2006 was followed.   First, the
``jitter'' files were repaired using the command {\tt cf\_jitter} and
the CalFUSE pipeline was run on the individual segments of the
observation.  Then, the command {\tt idf\_combine} was run to combine
the segments, setting the ``$-$c'' flag in order to recompute the
centroids of the spectral traces in the final intermediate data file
(IDF).  At the same time, the bad pixel maps were combined using {\tt
  bpm\_combine}.  These steps were run for data taken at night alone,
and for the combined day-and-night data.

\begin{figure}
\figurenum{14}
\epsscale{1.0}
\plotone{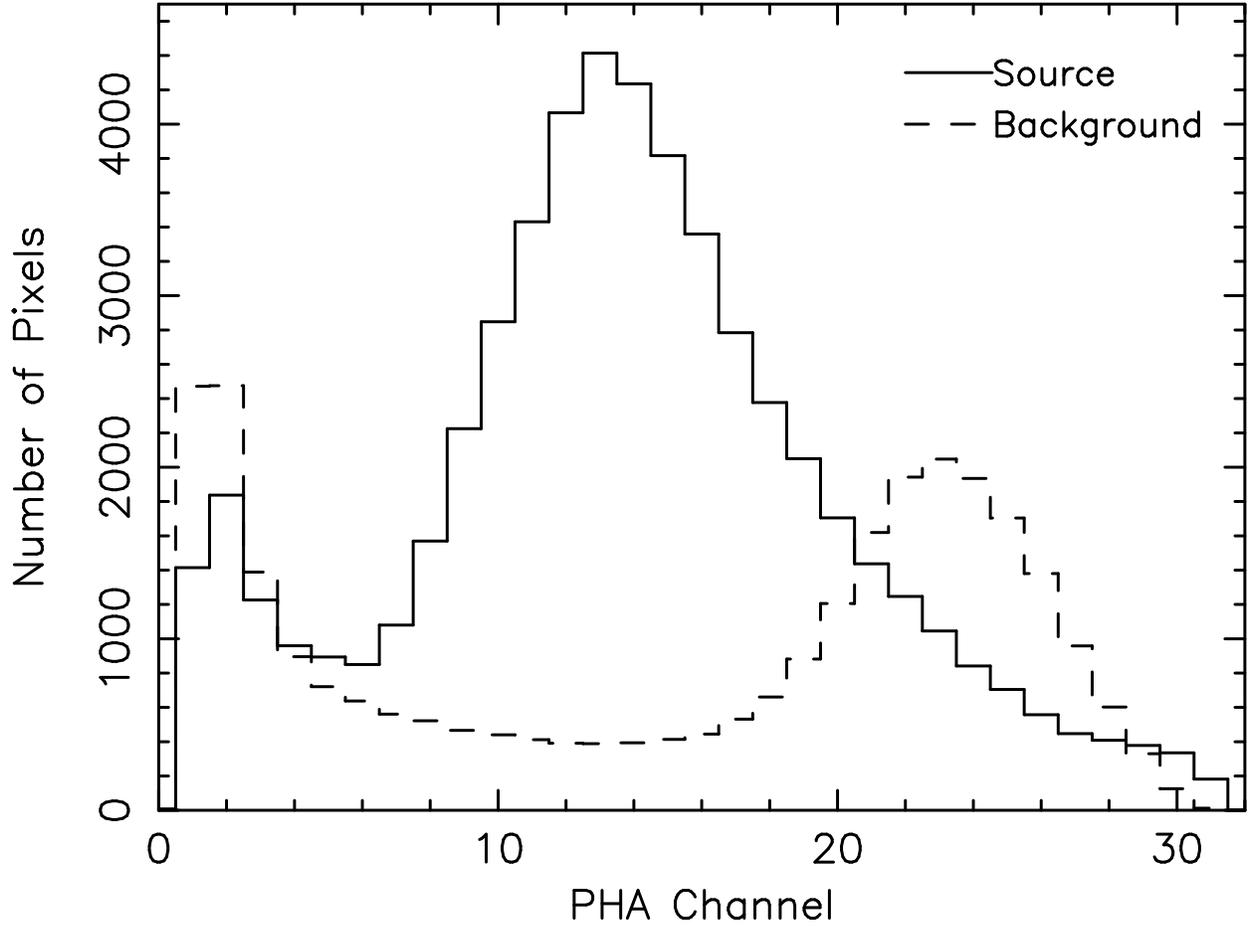}
\caption{Distribution of PHA from events lying in the source
 lif1a ``bowtie'' extraction region (solid line) and background
 regions scaled to the area of the source region (dashed line).\label{fig14}}
\end{figure}

At this point, the data are ready to have the spectrum extracted and
calibrated, and the background model constructed and the background
subtracted.  All of this is done by the CalFUSE script {\tt
  cf\_extract\_spectra}.  AGN are faint sources for 
{\it FUSE}, so a high-quality background subtraction is essential.
The data reduction and background subtraction pipeline for {\it
FUSE} has improved dramatically over the years \citep{dixon07}.
However, the background 
subtraction algorithms do not fully exploit an important fact: 
the background has a different PHA (pulse height analyzer) spectrum
than the source photons.  To illustrate this, we display in Fig.~\ref{fig14}
the PHA distributions for the source region and the background region
for the LIF1a data. Specifically, we extracted the data from the 
bow-tie LIF1a extraction region found in the spex1a009.fit file given
in CalFUSE version 2.4 (the extraction regions are handled differently
in CalFUSE 3.1 and greater), and the histogram of PHA is shown as a
solid line.  The  dashed line shows the PHA distribution from
background regions of the detector that avoid the apertures and
airglow lines, scaled to area of the source aperture, and therefore
approximately giving the distribution of background in the source
aperture.  This background is an estimate for two reasons: 1.) the
scattered light is more intense in the aperture, a fact that would
increase the estimation of the background; 2.) the regions of the
detectors under the apertures have lost sensitivity over the years, a
fact that would decrease the estimation of the background.  At any
rate, it can clearly be seen that the background dominates for PHA
channels less than 2 and greater than 20, and we can reduce the
background and therefore increase the statistics of the spectrum by
excluding these PHA channels.  Note that we are not the first to reduce
the background by imposing a PHA restriction
\citep[e.g.,][]{brotherton02,clb06}.    

The background files provided by the {\it FUSE} team have been
carefully constructed. Each one is comprised of $\sim 20$ background
observations, and they vary stepwise in time to account for periodically
imposed gain changes.  But they have been constructed for the default PHA 
range, and although they are scaled in the background modeling
process, in principle the shape of the background spectrum should
change depending on the PHA range.

We decided therefore to construct our own background images using the
following restricted PHA ranges:  1a: 5--20; 1b: 9--22; 2a: 3--14; 2b:
6--20.   These were determined as shown in Fig.~\ref{fig14}; we plotted
background and source regions, and determined the range of PHA in
which the source dominates.  To construct the background images, first
we determined the appropriate background observations from the lists
in the headers of the background files produced by the {\it FUSE}
team.  For our observation, there were 24 separate observation IDs
that were used in constructing the background images. In general, the
background is different during orbital day and orbital night; the
scattered background is a little higher in the day, plus there are
more and stronger airglow lines, and therefore we construct separate
background images for day and night.   These various observation
segments ranged from 765 seconds to 23345 seconds in duration; we
excluded the very lowest exposure observation from the 2a and 2b
detectors, therefore we used only 23 files to construct the
backgrounds for each of those detectors.    

For each detector and each segment, the following processing steps
were done.  The day and night intermediate data files were extracted
using the {\it FUSE} IDL program {\tt cf\_edit}, and at the same time
the pha restriction was applied.  In Calfuse 3.1 and higher, all
events are retained in the data files, and bad data (e.g., from
flaring or jitter, i.e., when the target is not in the aperture) are
marked using flags.  We extracted the flag information from the fits files,
and then use that information to separate out the good day or night
events. An image can be plotted using the good events.  

The image of each background file contains airglow lines, with the day
image having more lines and more intense lines.  We do not want the
intense lines in the background, so we exclude regions of the
image with strong lines.  The flags aid this selection as there is a
flag for airglow feature.  A mask is made of the excluded regions.  
Next, the extracted images are added together, day and night
separately, weighted by the exposure time and the airglow mask.
Regions with a fractional exposure less than 0.3 of the total exposure 
time are excluded from the background region.  The day and night
background images are convolved with a gaussian kernel with a FWHM of
15 pixels, and a constant is subtracted.  Finally, the images are
output as fits files in the same format as those produced by the {\it
  FUSE} team (indeed with the same headers, with exposure times
altered).  These files are available
online\footnote{http://www.nhn.ou.edu/~leighly/FUSE\_bkgd}. 

These background files can be used in two ways.   First, they can be
directly used in the CalFUSE script {\tt cf\_spec\_extract} by copying
them to the calfiles directory, and changing the BKGD\_CAL keyword in
the primary header of the IDF file from which the spectrum will be
extracted.   At the same time, PHALOW and PHAHIGH keywords should be
changed to match those of the background files.   The scaling factors
for the day and night data, plus the constant, can be extracted from
the verbose output of {\tt cf\_spec\_extract}.

\begin{figure}
\figurenum{15}
\epsscale{1.0}
\plotone{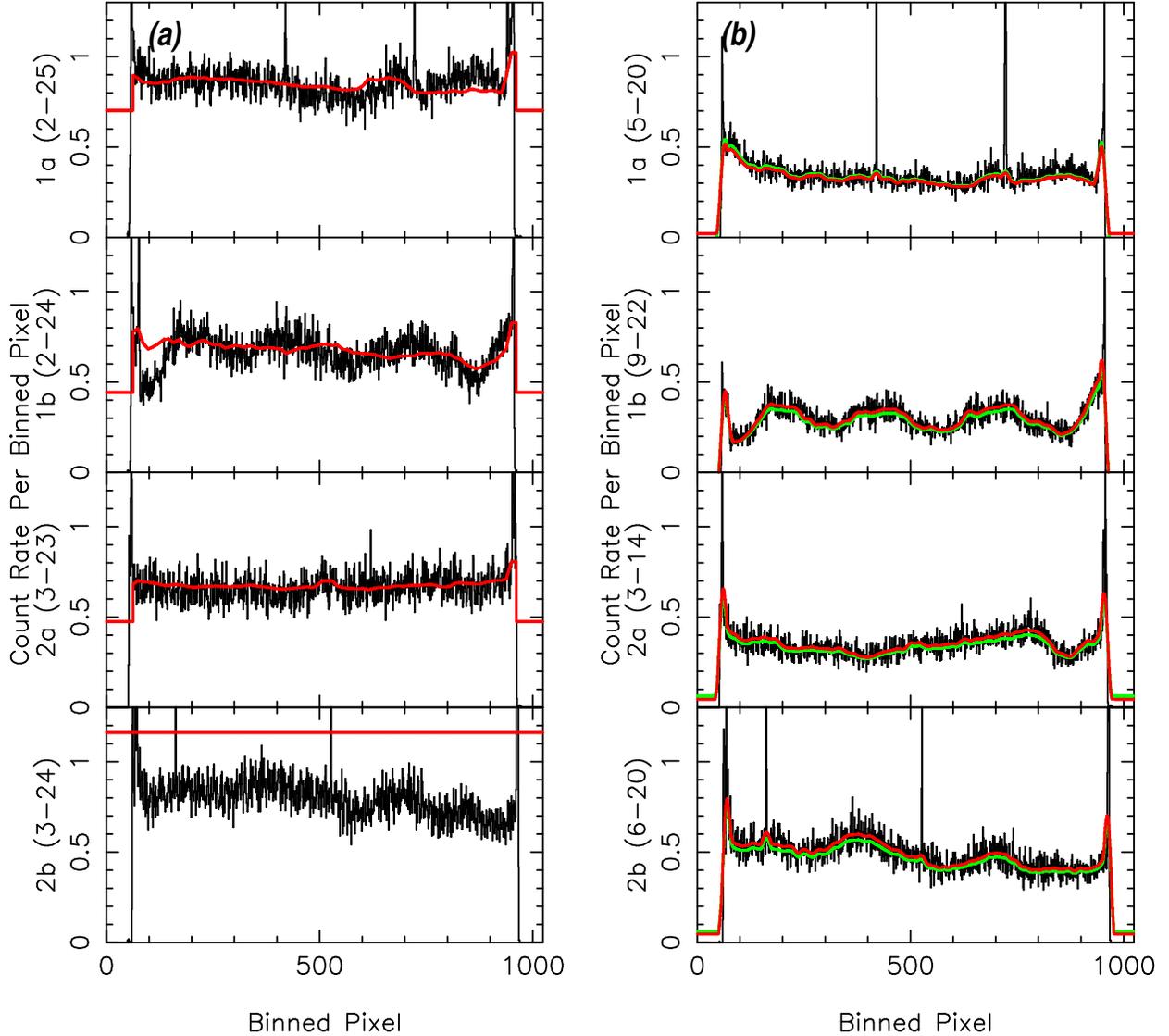}
\caption{Background spectra were created from the regions of the
  detectors avoiding the LWRS and MDRS apertures and high background
  regions near the edges of the detectors.  Background spectra were
  created from the background images over the same background
  regions.  Those background regions were scaled using either the {\tt
  cf\_extract\_spectra} output (red lines) or using the results from
  {\tt Specfit} modeling (green line; see text for details).  {\it
  (a.)} the default spectra and pha selection; (b.) the spectra from
  the new background images created by us using pha selection.  The
  background is lower when the PHA is restricted in range, and the
  background spectra model the data better.\label{fig15}}
\end{figure}

Fig.~\ref{fig15} shows the results of our background subtraction
compared with the default.  To construct these figures, we first
determined the y-direction ranges on the image in which the light from
the target or scattered emission is not seen, i.e., we avoid the LWRS
apertures and mostly the MDRS aperture as well.  In addition, regions
at the edge of the detector where the background is exceptionally high
were also avoided.  ``Spectra'' were constructed by summing the events
along the spatial direction, and binning by a factor of 16 in the
dispersion direction.  These spectra were extracted from the day+night
data from WPVS~007 (black lines) and from the modeled background data,
using the scale factors obtained from the {\tt cf\_spec\_extract}
output (red lines).  It can be readily seen that the
background-modeled spectra from our files match the data better. In
fact, in the case of the 2b detector, the {\tt cf\_spec\_extract}
modeling appears to have failed for the default data, as the day and
night scaling factors were zero, and the background was a constant;
the background modeling did not fail for these data for our background
files.

While Fig.~\ref{fig15} shows that the PHA-limited backgrounds perform
perfectly well in {\tt cf\_spec\_extract}, there is another way to use
these background files.  The background spectra were extracted from
the WPVS~007 data as above, and day and night background spectra were
extracted from the model files over the same background regions.
Then, the WPVS~007 background spectra were fit using the IRAF task
{\tt Specfit} to a model consisting of the day background spectrum,
the night background spectrum, and a constant.  Regions where the flux
is zero (i.e., beyond the edges of the detector) and regions where
strong airglow lines are present were excluded from the fit. The
normalizations of the day and night files were constrained to be in
proportion to their respective exposure times.  Then, the {\tt
  Specfit} results could be input directly into a modified version of
the {\tt cf\_spec\_extract} program.  The result is shown in the right
panel as the green line in Fig.~\ref{fig15}.  It differs only slightly
from the result obtained using {\tt cf\_spec\_extract} directly.

Fig.~\ref{fig16} shows the a comparison of the LIF spectra extracted
using the default PHA ranges and {\tt cf\_spec\_extract} program with
spectra obtained using the restricted PHA range and the new background
files scaled using the {\tt Specfit} fitting described above.  We show
both the day-and-night spectra in dark grey, and night-only spectra in
light grey.  The chief utility of the night-only spectra is the
identification of airglow lines in the day-and-night spectra.  In
addition, the less prominent airglow lines are seen only in the day
data, so they can be removed by substituting the night-only spectra in
those regions.  Overall, there is not much difference between the
default and the PHA-restricted spectra, even though the background
levels are 5--40\% lower for the PHA-restricted spectra.  There are
detailed differences, however. For example, the 1b day-and-night
spectra do not agree in normalization with the night-only spectra for
the default reduction, while they do for the PHA-restricted reduction.
In addition, the failure of the background subtraction for the
day-and-night Lif2b data is apparent for the default reduction.

Fig.~\ref{fig17} shows a similar plot for the SIC data.  In this case,
only the night spectra are shown.  The SIC has a much lower effective
area (a factor of three) than the LIF, and it is difficult to obtain
useful results from these detectors for a faint object like WPVS~007.
This can be seen by comparing the spectrum around 1065\AA\/ in the
LIF1a spectrum in Fig.~\ref{fig16}, and the same region in the SIC1a
spectrum.  However, the SIC spectra extend to shorter wavelengths; in
particular, a feature is clearly detected near 980\AA\/ in both the
SIC1b and SIC2a spectra. Again, overall, the differences between the
default spectra and the restricted-PHA spectra are small; however, the
restricted PHA spectra are slightly less noisy, and approach zero more
gracefully (e.g., SIC2a for wavelengths shorter than 960\AA\/).

\section{Merging the {\it FUSE} Spectra}

The {\it FUSE} observation yields eight separate spectra, and we
proceed here to merge them.  WPVS~007 is a faint object for {\it
  FUSE} and as the SIC telescope/detector systems have effective areas
about 1/3 of the LIF, the SIC spectra have rather low signal-to-noise
ratios (Fig.~\ref{fig17}).  Thus, we use those data only for the shortest
wavelength ranges where LIF spectra are not available.  

\begin{figure}
\figurenum{16}
\epsscale{1.0}
\plotone{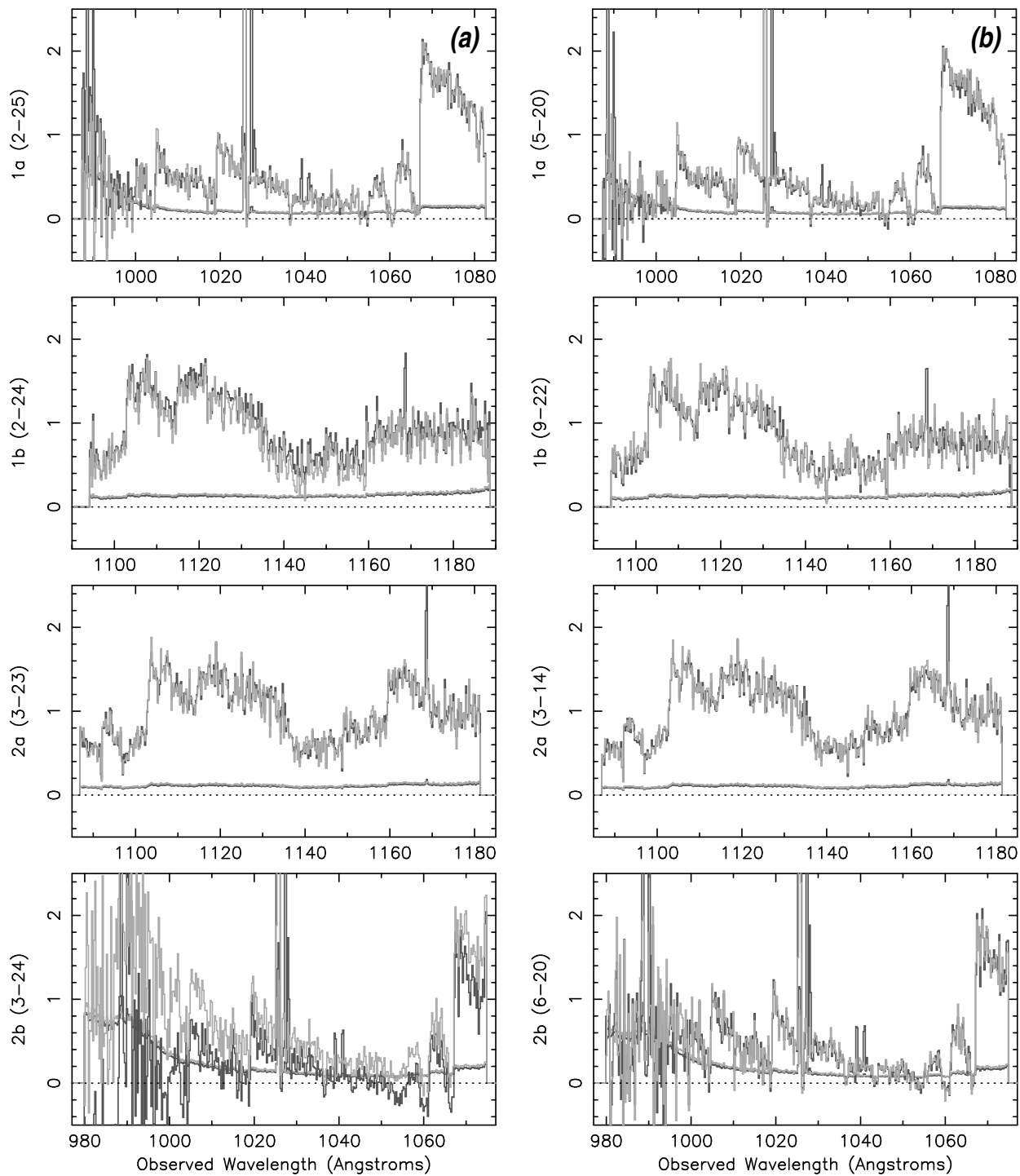}
\caption{Comparison of LIF spectra extracted using default PHA ranges
  and background images (left, (a)) with spectra extracted using the
  restricted PHA range and the new background images (right, (b)).
  Day-and-night spectra are shown in dark grey, and night-only
  spectra are shown in light grey.    In both cases, the uncertainty is
  shown as a line below the spectra for clarity.  These figures show
  that overall there is not very much difference between the default
  and new spectra.\label{fig16}}
\end{figure}

\begin{figure}
\figurenum{17}
\epsscale{1.0}
\plotone{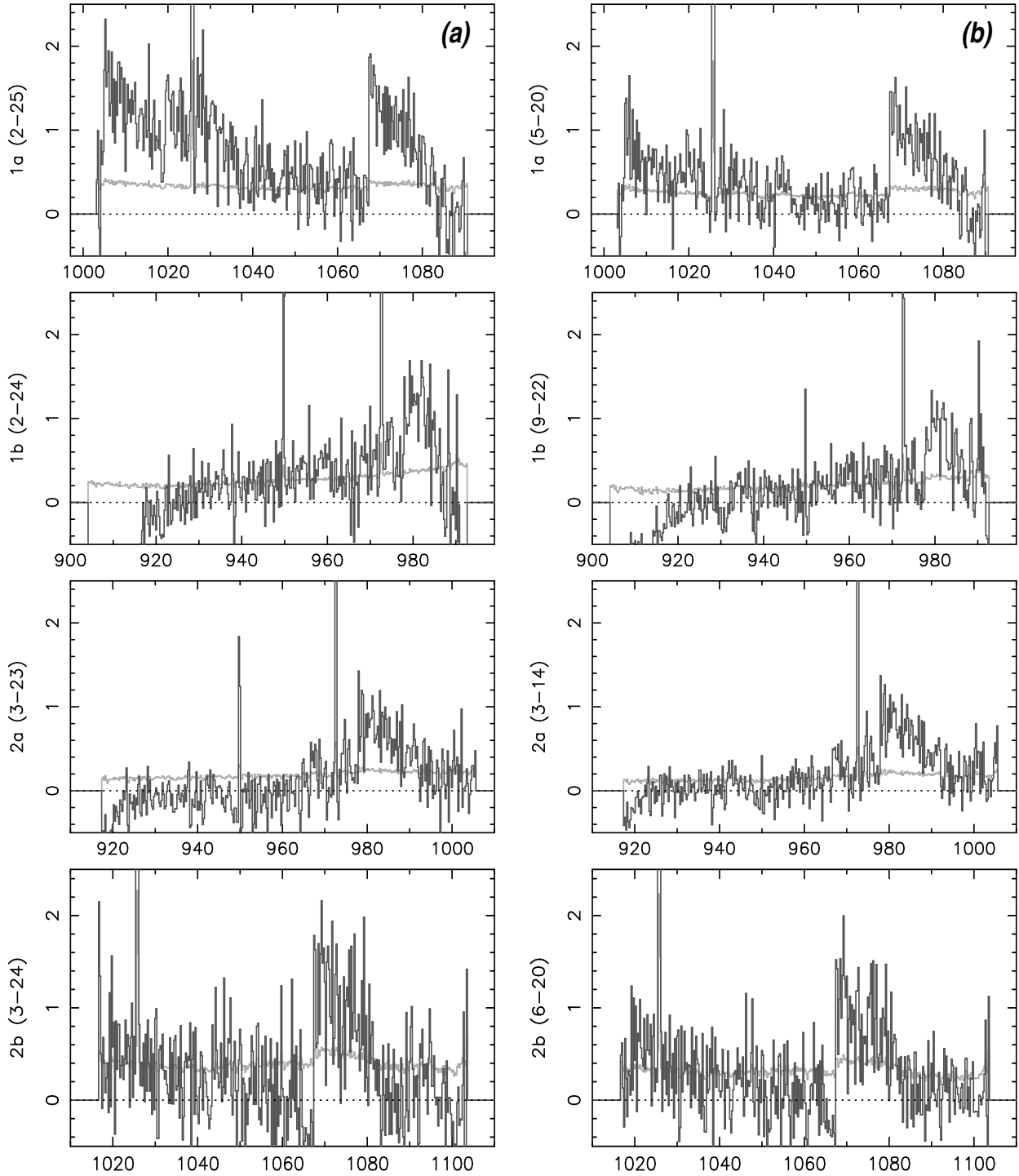}
\caption{Similar to Fig.~\ref{fig16} for the SIC spectra.  The left panel (a)
  shows the spectra extracted using default PHA ranges and background
  images; the right panel (b) shows the spectra extracted with the
  restricted PHA range and new background.  In both cases, night-only
  spectra are shown.\label{fig17}}
\end{figure}

The longest wavelengths are sampled by the LIF 1b and LIF 2a.  LIF 1b
is commonly afflicted by the ``worm'', or shadowing of the grid wires
on the detector \citep{sahnow02}.  Our spectra are no exception, as
can be seen by the difference in the 1b and 2a spectra longward of
$\sim 1140$ \AA\/ (Fig.~\ref{fig16}).  The worm decreases the
effective area, decreasing also the signal-to-noise ratio, and
therefore we opt to ignore afflicted region.  We average the two
spectra in overlapping region (1094.25--1137.25 \AA\/), and use the 2a
spectrum longward of that.  The errors are computed by propagation in
quadrature.

We turn next to the region of the spectrum between 987.5 and 1074.75
\AA\/, which are represented by both the LIF 1a and LIF 2b spectra.
The LIF 2b spectrum has generally a poorer signal-to-noise ratio than
the LIF 1a spectrum, as seen in Fig.~\ref{fig16}, and the question
arises, do we enhance or degrade the signal-to-noise ratio of the
LIF1a spectrum by averaging it with the LIF2b spectrum?  We also find
that the normalization of the 2b spectrum appears to be about 15\%
lower than the 1a spectrum.  Since at the time of this observation,
the pointing was still being determined using the LIF1a detector, we
assume that it has the correct normalization.  We examine the  mean
spectrum and decide to use the mean of the LIF1a and LIF2b scaled by a
factor of 1.15 between 1011 and 1074.75 \AA\/.  The LIF1a contains
useful information longward of the merged region, to $\sim 1082.5$
\AA\/, and there is a gap between the coverage of LIF1a and the
coverage of LIF2b from $\sim 1082.5$ to $\sim 1087$\AA\/.  In
principle,  the SIC 2b could be used to fill this gap; however,
examination of Fig.~\ref{fig17} shows there is no signal in this
region in that spectrum. 

Shortward of 1100\AA\/, we use the LIF1a, SIC1b, and SIC2a spectra, as
follows.  The LIF1a nominally extends down to 987.5 \AA\/, but as can
be seen in Fig.~\ref{fig16}, the signal-to-noise ratio approaches 1 at the
shortest wavelengths.  The SIC2a overlaps up to 1005.5\AA\/, but
again, the errors are large at the end of the spectrum.  We use Lif1a
alone down to 1002.75\AA\/, and use the mean of Lif1a and Sic2a
between 995.5 and 1002.5\AA\/.  Shortward of that, the spectrum is
represented by Sic2a until 992.5\AA\/, when it is joined by Sic1b.
The uncertainty on the Sic1b spectrum is very large at the end, so we
use the Sic2a alone down to 987.75\AA\/.  Shortward of that, we use
the mean of the Sic1b and Sic2a spectra down to 920\AA\/.  

There are several prominent airglow lines remaining in the spectrum.
These include the Lyman lines of hydrogen, Ly$\beta$ at 1025.722\AA\/,
Ly$\gamma$ at 972.537\AA\/, and Ly$\delta$ at 949.743\AA\/, and
\ion{O}{1} line near 988\AA\/.  The regions of the spectra in the
vicinity of these lines are excised.  

Finally, the spectra are modestly smoothed\footnote{The smoothing
function is $0.2(f(i-1)+3f(i)+f(i+1))$.}, dereddened and shifted to
the restframe.

\clearpage

\begin{deluxetable}{lccccc}
\tablecaption{Observing Log}
\tablewidth{0pt}
\tablehead{
\colhead{Spectrometer} & \colhead{Date} & \colhead{Exposure} &
\colhead{Bandpass} & \colhead{Aperture\tablenotemark{a}} &
\colhead{Resolution} \\
&& \colhead{(seconds)} & \colhead{(\AA\/)} && \colhead{(\AA\/)}}
\startdata
{\it HST} FOS (G130H) & 1996-07-30 & 3840 & 1140--1606 &
$0.86^{\prime\prime}$ & 2.26 \\
{\it HST} FOS (G190H) & 1996-07-30 & 1500 & 1590--2312 & 
$0.86^{\prime\prime}$ & 3.16 \\
{\it HST} FOS (G1270H) & 1996-07-30 & 1280 & 2222--3277 &
$0.86^{\prime\prime}$ & 4.72 \\
{\it HST} FOS (G1400H) & 1996-07-30 & 1000 & 3235--4781 &
$0.86^{\prime\prime}$ & 6.94 \\
{\it HST} FOS (G570H) & 1996-07-30 & 630 & 4569--6818 &
$0.86^{\prime\prime}$ & 10.06 \\
{\it FUSE} LIF1a & 2003-11-06 & 47354 & 987.5--1082.5 &
$30^{\prime\prime}$ & 0.25\tablenotemark{b} \\
{\it FUSE} LIF1b & 2003-11-06 & 48054 & 1094.25--1188.75 & 
$30^{\prime\prime}$ & 0.25\tablenotemark{b} \\
{\it FUSE} LIF2a & 2003-11-06 & 48229 &  1087.0--1181.25 & 
$30^{\prime\prime}$ & 0.25\tablenotemark{b} \\
{\it FUSE} LIF2b & 2003-11-06 & 48466 & 980.0--1074.75 & 
$30^{\prime\prime}$ & 0.25\tablenotemark{b} \\
{\it FUSE} SIC1a & 2003-11-06 & 35055 & 1003.25--1090.5 & 
$30^{\prime\prime}$ & 0.25\tablenotemark{b} \\
{\it FUSE} SIC1b & 2003-11-06 & 35259 & 904.25--992.5 & 
$30^{\prime\prime}$ & 0.25\tablenotemark{b} \\
{\it FUSE} SIC2a & 2003-11-06 & 34982 & 917.5--1005.5 & 
$30^{\prime\prime}$ & 0.25\tablenotemark{b} \\
{\it FUSE} SIC2b & 2003-11-06 & 35010 & 1016.75--1103.5 & 
$30^{\prime\prime}$ & 0.25\tablenotemark{b} \\
\enddata
\tablenotetext{a}{For the {\it HST} FOS spectra, this refers
  to the size of the round aperture.  For the {\it FUSE} spectra,
  this refers to the size of the LWRS square aperture.}
\tablenotetext{b}{The observed resolution of {\it FUSE} using the LWRS
  aperture is $R=20,000 \pm 2000$ (The FUSE Observer's
      Guide: http://fuse.pha.jhu.edu/support/guide/guide.html).
  However, the low count rate cannot sample this resolution.
  Therefore, the resolution refers to the final binsize of the
  spectra.}
\end{deluxetable}

\newpage

\begin{deluxetable}{lrccc}
\tablecaption{Estimated Column Densities}
\tablewidth{0pt}
\tablehead{
\colhead{Ion} & \colhead{Wavelength} & \colhead{Scale Factor} &
\colhead{Template\tablenotemark{a}} & \colhead{$\log(N)$\tablenotemark{b}}  \\
& \colhead{(\AA\/)} & & &  \colhead{($\rm cm^{-2}$)} 
}
\startdata
\ion{P}{4} &  950.66 &  $3.08 \pm 0.38$ &   \ion{P}{5} & $>15.19 \pm 0.05$ \\
\ion{P}{4} &  950.66 & $1.33 \pm  0.58$ &  \ion{O}{6} MiniBAL & $>14.31 \pm 0.20$ \\

Ly$\gamma$\tablenotemark{c} &  972.54 & $2.17\pm  0.31$ & \ion{P}{5} &
$>16.77 \pm  0.06$  \\

\ion{C}{3} &  977.03 & $2.61 \pm 0.37$ &  \ion{P}{5} & $>15.43 \pm  0.06$
\\
\ion{C}{3} &  977.03 & $1.28 \pm 0.23$  & \ion{O}{6} MiniBAL & $>14.60 \pm  0.08$  \\
\ion{C}{3}\tablenotemark{d} &  977.03 & 1 &  & $>15.87$ \\

\ion{N}{3}\tablenotemark{c} &   990.98 & $2.38\pm 0.09$ &  \ion{P}{5}
& $>15.87 \pm  0.02$ \\
\ion{N}{3}\tablenotemark{c} &  990.98 & $0.88 \pm 0.23$ &  \ion{O}{6}
MiniBAL & $>14.93 \pm  0.12$  \\

\ion{P}{3} & 998.00 & $<0.27$ & \ion{P}{5} & $<15.26$  \\
\ion{P}{3} & 1001.73 & $0.59\pm 0.25$ & \ion{P}{5} & $>15.60 \pm 0.19$ \\ 
\ion{P}{3} & 1003.60 & $0.77\pm 0.29$ &  \ion{P}{5} & $>15.72 \pm 0.17$ \\

\ion{S}{3}  &  1012.50 & $1.10 \pm 0.34$ &  \ion{P}{5} &$>16.29 \pm 0.14$ \\
\ion{S}{3}\tablenotemark{c}  &  1015.63 & $1.26 \pm 0.47$ &
\ion{P}{5} &  $>16.35 \pm 0.17$  \\
\ion{S}{3}\tablenotemark{c}  &  1021.32 & $1.28\pm  0.46$ & \ion{P}{5}
& $>16.35 \pm 0.16$ \\ 

Ly$\beta$\tablenotemark{c} & 1025.72 & $2.72 \pm 0.46$ &  \ion{P}{5} &
$>16.41 \pm  0.07$  \\
Ly$\beta$\tablenotemark{c} & 1025.72 & $3.89 \pm 3.0$ &  \ion{O}{6}
MiniBAL & $>16.05 \pm  0.44$ \\

\ion{O}{6} & 1031.91 & $1.68 \pm 0.28$ &  \ion{P}{5} & $>15.97 \pm 0.07$ \\
\ion{O}{6} & 1037.61 & $2.41 \pm 0.15$ &  \ion{P}{5} & $>16.43 \pm 0.03$ \\

\ion{O}{6} & 1031.91 & $3.0 \pm 1.1$ &   \ion{O}{6} MiniBAL & $>15.71 \pm 0.17$  \\
\ion{O}{6} & 1037.61 & $1.99 \pm  0.25 $ &  \ion{O}{6} MiniBAL & $>15.83 \pm 0.06$ \\

\ion{O}{6}\tablenotemark{d} & 1033.68 & 1 &  &  $>16.64$ \\

\ion{S}{4} & 1062.66 & $1.20 \pm 0.05$ &  \ion{P}{5} &  $>15.17 \pm 0.02$  \\
\ion{S}{4}\tablenotemark{c} & 1073.03 & $1.54 \pm 0.03$ &  \ion{P}{5}
& $>15.27 \pm  0.01$  \\
\ion{S}{4}\tablenotemark{c,e} & 1073.03 & $<0.046$ &  \ion{O}{6}
MiniBAL &   $<13.24$ \\ 

\ion{N}{2}\tablenotemark{c,e} & 1085.12 & $0.09 \pm 0.02$ &
\ion{P}{5} &  $>13.77 \pm 0.11$ \\

\ion{P}{5} & 1117.98 & $1.30 \pm  0.04$ &  \ion{P}{5} & $>15.29 \pm  0.01$  \\
\ion{P}{5} & 1128.01 & $0.89 \pm  0.04$ &  \ion{P}{5} & $>15.43 \pm  0.02$ \\
\ion{P}{5}\tablenotemark{e} & 1128.01 & $<0.09$ &  \ion{O}{6} MiniBAL
&  $<13.95$ \\

\ion{Fe}{3}\tablenotemark{c,e} & 1125.79 & $<0.17$ &  \ion{P}{5} & $<13.77$ \\

Ly$\alpha$\tablenotemark{c} & 1215.67 & $0.95 \pm 0.05$ & \ion{N}{5}
MiniBAL &  $>12.31 \pm 0.02$ \\
\ion{N}{5} & 1238.82 & $1.07 \pm  0.13$ &  \ion{N}{5} MiniBAL& $>12.78 \pm 0.05$  \\
\ion{N}{5} & 1242.80 & $0.94 \pm  0.11$ & \ion{N}{5} MiniBAL &  $>13.02 \pm 0.05$  \\
\ion{C}{4} & 1548.20 & $1.27 \pm  0.10$ &  \ion{N}{5} MiniBAL & $>12.67 \pm 0.03$ \\
\ion{C}{4} & 1550.77 & $1.59 \pm  0.09$ &  \ion{N}{5} MiniBAL  & $>13.05
\pm 0.03$ \\
\enddata
\tablenotetext{a}{Development of the three templates are described in
the following sections: \ion{P}{5} in \S 3.1; \ion{O}{6} MiniBAL in \S
3.2; \ion{N}{5} MiniBAL in \S 2.3.}
\tablenotetext{b}{The ionic column densities were obtained from the
apparent optical depths and therefore are lower limits.  The detections
are denoted using $>$, and the uncertainties reflect the statistical
uncertainty on the determination of the lower limit in the spectral
fitting.}
\tablenotetext{c}{When the line is comprised of multiplets
  indistinguishable at the spectral resolution, the column densities
  were estimated using the sum of the $f_{12}$ for the multiplets.}
\tablenotetext{d}{Column density obtained using the alternative
  deblending described in \S 3.3.4.  No template is used.}
\tablenotetext{e}{Upper limits on lines that were not observed.}
\end{deluxetable}

\newpage

\begin{deluxetable}{lcccccccc}
\rotate
\tabletypesize{\scriptsize}
\tablecaption{Estimated X-ray 0.3--10 keV Count Rates}
\tablewidth{0pt}
\tablehead{
\colhead{Model} & \multicolumn{4}{c}{{\it Chandra}} &
\multicolumn{4}{c}{\it Swift} \\
& \multicolumn{4}{c}{\hrulefill} & \multicolumn{4}{c}{\hrulefill} \\
& \colhead{direct rate} & \colhead{direct counts\tablenotemark{a}} & \colhead{absorbed
  rate} & \colhead{absorbed counts\tablenotemark{a}} & 
\colhead{direct rate} & \colhead{direct counts\tablenotemark{b}} & \colhead{absorbed
  rate} & \colhead{absorbed counts\tablenotemark{b}}
}
\startdata
$\alpha_{ox}=-1.28$; $N_H=23.0$ & 1.53 & 14192 &  0.049 & 454 & 0.15 &  12599 &
0.0038 & 325 \\
$\alpha_{ox}=-1.9$; $N_H=22.2$ &  0.032 &  300 & 0.0061 &  56 &  0.0090 & 764 &
0.0011 & 91 \\
$\alpha_{ox}=-1.9$, $Z=5$;  $N_H=21.6$ & 0.032 & 300 & 0.0045 &  41 &  0.0089 &
764 & 0.00083 & 71 \\
\enddata
\tablenotetext{a}{The effective exposure of the {\it Chandra}
  observation was 9300 seconds, and 10 net source photons were observed.}
\tablenotetext{b}{The effective exposure of the {\it Swift}
  observation was 85,508 seconds \citep{grupe08}, and 35.7 net source
  photons were observed.}
\end{deluxetable}

\newpage

\begin{deluxetable}{lcccccc}
\tablecaption{Dynamics Components}
\tablewidth{0pt}
\tablehead{\colhead{Model} & \colhead{$L_{46}$\tablenotemark{a}} &
  \colhead{$L/L_{Edd}$\tablenotemark{b}} &  \colhead{$\rm  F_{0.1}$\tablenotemark{c}}&
\colhead{$\rm N_{22}$\tablenotemark{d}} & \colhead{$\rm
  R_{0.1}$\tablenotemark{e}} \\
}
\startdata
$\alpha_{ox}=-1.28$ & 0.0114 & 0.22 &  4.1  & 10.0 &  0.0031 \\
$\alpha_{ox}=-1.9$ & 0.0049 & 0.096 & 0.67 & 1.58 & $<0$\\
$\alpha_{ox}=-1.9$, $Z=5$ & 0.0049 & 0.096 &  0.44 & 0.40 &  0.0070 \\
\enddata
\tablecomments{We use a black hole mass $M_8=0.041$ estimated from the
  optical luminosity and the H$\beta$ line width \citep{collin06}.  We
  use a terminal velocity $V_\infty=4000 \rm \, km\, s^{-1}$ estimated from the apparent
  optical depth shown in Fig.~\ref{fig4}.}

\tablenotetext{a}{The bolometric luminosity  in units of
  $10^{46}\rm \, erg\, s^{-1}$ obtained by integrating
  over the {\it Cloudy} input continua discussed in \S 4.1.1 after
  normalizing to the {\it HST} spectrum in the optical band.}
\tablenotetext{b}{The ratio of the bolometric luminosity to the the
  Eddington luminosity obtained using the black hole mass estimate $M_8=0.041$.}
\tablenotetext{c}{The fraction of the continuum absorbed by the wind,
  estimated by integrating over the absorbed continuum, and divided by
0.1.  }
\tablenotetext{d}{The lower limit on the column densities obtained in
  \S 4.1.2 in units of $10^{22}\rm \, cm^{-2}$.  }
\tablenotetext{e}{The launch radius in units of $0.1 \rm \, pc$
  estimated using Eq.\ 3 in    \citet{hamann98}.}
\end{deluxetable}

\vfill\eject



\clearpage

\clearpage

\clearpage

\clearpage

\clearpage

\clearpage

\clearpage

\clearpage

\clearpage

\clearpage

\clearpage

\clearpage

\clearpage

\clearpage

\clearpage

\clearpage

\clearpage








\clearpage


\end{document}